\documentstyle[epsf,times]{mn}
\begin{document}

\title[Universal gas density and temperature profile]
 {Universal gas density and temperature profile}

\author[E. Komatsu \& U. Seljak]
 {E.~Komatsu$^{1,2}$\thanks{E-mail: komatsu@astro.princeton.edu} 
 and U.~Seljak$^3$\thanks{E-mail: uros@feynman.princeton.edu} \\
 $^1$Department of Astrophysical Sciences, Princeton University,
 Princeton, NJ 08544, USA\\
 $^2$Astronomical Institute, T\^ohoku University, Aoba, Sendai 980-8578,
 Japan\\
 $^3$Department of Physics, Princeton University, Princeton, NJ 08544,
 USA}

%\date{}

\pagerange{\pageref{firstpage}--\pageref{lastpage}}
\pubyear{2001}

\maketitle

\label{firstpage}

%%%%%%%%%%%%%%%%%%%%%%%%%%%%%%%%%%%%%%%%%%%%%%%%%%%%%%%%%%%%%%%%%%%
\begin{abstract}
 We present an analytic approach to predict gas density and 
 temperature profiles in dark matter haloes.
 We assume that the gas density profile traces the dark matter density
 profile in outer parts of the haloes, as suggested by many 
 hydrodynamic simulations. 
 Under this assumption, the hydrostatic equilibrium uniquely fixes 
 the two free parameters in the model, the mass--temperature normalization and 
 the polytropic index, that determine the temperature profile.
 This enables us to predict a universal gas profile from any universal 
 dark matter profile.
 Our results show that gas tracing dark matter in the outer parts of the 
 haloes is inconsistent with gas being isothermal; on the contrary,
 it requires temperature to decrease slowly with radius, in agreement with 
 observations. 
 We compare our predictions for X-ray surface brightness profiles of the
 haloes and the mass--temperature relation with observations. 
 We find that they are generally in a good agreement.
 We compare the universal profile with the $\beta$ profile and find that,
 although the $\beta$ profile gives a reasonable fit to our predicted 
 profiles, the deviation from it can explain many of the observed trends,
 once we take into account the observational selection effects. 
 Our model predicts that the mass--temperature relation does
 not follow the self-similar relation because of the mass-dependent 
 halo concentration. 
 We also predict surface brightness profiles of the Sunyaev--Zel'dovich 
 (SZ) effect.
 We find that fitted to the $\beta$-profile the core radii and 
 $\beta$ inferred from the SZ effect are systematically larger than 
 those from the X-ray measurement.
\end{abstract}
%%%%%%%%%%%%%%%%%%%%%%%%%%%%%%%%%%%%%%%%%%%%%%%%%%%%%%%%%%%%%%%%%%%
%%%%%%%%%%%%%%%%%%%%%%%%%%%%%%%%%%%%%%%%%%%%%%%%%%%%%%%%%%%%%%%%%%%

%%%%%%%%%%%%%%%%%%%%%%%%%%%%%%%%%%%%%%%%%%%%%%%%%%%%%%%%%%%%%%%%%%%
\section{Introduction}

%%%%% motivation to investigate small scale structures %%%%%

The structure of the dark matter haloes has been 
recognized as one of the most powerful probes of small scale physics, 
which may potentially distinguish between the competing theoretical models. 
In cold dark matter (CDM) models, the structure forms through a sequence of 
mergers, which gives a characteristic density profile of the 
resulting haloes. 

Numerical simulations led the way in our understanding of the structure 
of the haloes. 
While there is significant scatter among the individual simulated haloes, 
one finds some similarity among them, justifying the introduction of 
a universal dark matter halo profile. 
The density of such a profile changes steeply in the outer parts of
haloes, scaling roughly as $\rho \propto r^{-\alpha}$ with $\alpha \sim 3$.
The slope gradually becomes shallower in the inner parts, converging to 
a value between $1<\alpha<3/2$ (Navarro, Frenk \& White 1996, 1997; 
Moore et al. 1998).
The transition between the two regimes is in general mass dependent, 
because the lower mass haloes collapse at a higher redshift, 
when the mean density of the universe was higher. 
This in turn requires a higher central density of the lower mass haloes.

Different dark matter models make different predictions for the slope 
of the inner profile as well as for the value of 
the central density as a function of the halo mass: 
both CDM and warm dark matter (WDM) 
predict cuspy haloes, but CDM in general predicts 
denser haloes at a given mass and radius. 
Collisional dark matter, however, predicts a central core.

%%%%% universal gas density and temperature profiles %%%%%

While the dark matter density profiles are important, they cannot be easily observed. 
In this paper, we focus instead on the gas density and temperature profiles, 
which can be observed through the X-ray emission and the 
Sunyaev--Zel'dovich (SZ) effect.
We introduce a concept of a universal gas density profile
in analogy with which an universal dark matter profile
describes, in an average sense, how the gas density depends on radii. 
Such a construction is as useful as the universal dark matter density profile, 
in the sense that it can parameterize, in a simple way, how different 
cosmological models differ in their predictions for the gas density profile,
and place constraints on the models from the observations. 

%%%%% limitation of the beta profile %%%%%

By far the most common form used for the gas density profile is the so-called 
$\beta$ profile, which consists of a central core and a power-law outer 
slope, $\beta$. 
While such a profile is useful in observational description of the
X-ray clusters, it becomes insufficient when bolometric SZ observations 
become available; these observations will measure the gas density 
profile out to the outer parts of the clusters, where the predicted 
slope of the profile changes and can no longer be described by a 
single power law. 

As we discuss in this paper, there are already 
some indications that the $\beta$ profile is
insufficient to describe the X-ray data. 
For example, we can explain a correlation between core radii and $\beta$, found
in observational data, as a consequence of forcing 
the $\beta$ profile to describe the underlying gas density profile
in which the slope increases continuously with radius. 

Another shortcoming of the $\beta$ profile is that it is not derived
from a dark matter density profile; thus, we cannot relate
the observed X-ray profile to the underlying dark matter profile
in simple terms. 
In this paper, we give a prescription of how to compute the 
gas density and temperature profiles from any given dark matter density 
profiles.
We show that the prescription is unique (no free parameters required) 
under our assumptions. 
This property causes our model to differ from similar recent 
constructions by Makino, Sasaki \& Suto (1998) and Suto, Sasaki \&
Makino (1998).

%%%%% on the assumptions %%%%%

To construct the gas density and temperature profiles, 
we assume that the gas is in hydrostatic equilibrium throughout the halo. 
This in itself does not suffice to construct a gas density profile, 
because the dependence of the temperature on radii remains unspecified. 
Assuming isothermal profile gives a poor fit to the 
observed clusters \cite{MSS98} and is not favored observationally 
(Finoguenov, Reiprich \& B\"ohringer 2001). 

Instead, we make an alternative assumption that the gas density profile 
traces the dark matter density profile in the outer parts of the halo. 
While this assumption is qualitatively reasonable, its quantitative 
justification comes from hydrodynamic simulations that show remarkable 
agreement between the two profiles in the outer parts of haloes. 
It is possible that some physics is missing in the current generation of
numerical simulations, which for the most part ignore radiative processes. 
Gas cooling and star formation can affect not only the central regions, but 
also the global properties of the cluster; however, even in that case, 
the gas profile in the outer regions remains relatively unaffected 
\cite{PTCE00,Lewis00}. 

The remarkable property is that once assuming that the gas profile 
traces the dark matter profile outside the core, and the gas 
obeys a polytropic equation of state, we can predict both the 
mass--temperature normalization and the temperature dependence on radii 
analytically. 
we find that the predictions agree with observations as well as 
simulations.

Our assumption that gas temperature is monotonic with density 
limits us to regions outside the core (roughly outside
inner $100-200$~kpc), because in the inner regions gas temperature is often 
increasing with radius up to $100-200$~kpc and then mildly 
decreasing in the outer regions. 
For the most part, however, we will compare our results with the 
observational data for which the central part of the cluster has been excised 
(e.g. Finoguenov et al. 2001).
We are thus justified to make this assumption; consequently, 
our model does not provide constraints on the gas and dark matter 
profiles in the inner parts of the cluster.

%%%%% cosmological parameters %%%%%

Although the cosmology does not play a major role, 
for the purpose of this paper, we fix the cosmological parameters at the
best-fitting values from CMB and large-scale structures \cite{Boom01}:
$\Omega_{\rm m}=0.3$, $\Omega_{\Lambda}=0.7$, $\Omega_{\rm b}=0.05$, 
$h=0.67$, $n=1$, and $\sigma_8=0.9$.
We evaluate all the predictions at the present epoch ($z=0$).

%%%%% structure of the paper %%%%%

The structure of this paper is as follows. 
In \S 2, we describe the universal dark matter haloes and relations between 
the virial mass and the observed mass. 
In \S 3, we introduce a family of self-similar gas density and temperature 
profiles.
We show how the assumptions fix free parameters in the model.
After discussing relations between the virial temperature and the observed gas
temperature, we predict gas density and temperature profiles of haloes
explicitly.
In \S 4, we compare the X-ray surface brightness profiles derived from
the predicted gas profiles with the $\beta$ profile.
We study the observational selection effect on the comparison in detail. 
In \S 5, we discuss the surface brightness profiles of the SZ effect. 
%and then discuss a possibility of testing our model observationally.
In \S 6, we predict the mass--temperature relation, and compare it with
observations. 
Finally, we summarize our results in \S 7. 

%%%%%%%%%%%%%%%%%%%%%%%%%%%%%%%%%%%%%%%%%%%%%%%%%%%%%%%%%%%%%%%%%%%
\section{Universal Dark Matter Density Profiles}

\subsection{Basic properties}

Many high-resolution $N$-body simulations suggest that the 
dark matter density profile, $\rho_{\rm dm}(r)$, is well described 
by a self-similar form. The profile is expressed with a 
characteristic radius $r_{\rm s}$:
%%%%%%%%%%%%%%%%%%%%%%%%%%%%%%%%%%%%%%%%%%%%%%%%%%%%%%%%%%%%%%%%%%%
\begin{equation}
 \label{eq:dmprofile}
  \rho_{\rm dm}(r)= \rho_{\rm s} y_{\rm dm}(r/r_{\rm s}),
\end{equation}
%%%%%%%%%%%%%%%%%%%%%%%%%%%%%%%%%%%%%%%%%%%%%%%%%%%%%%%%%%%%%%%%%%%
where $\rho_{\rm s}$ is a normalization factor in units of mass density, 
while $y_{\rm dm}(x)$ is a non-dimensional function representing the profile. 
We have chosen $\rho_{\rm s}$ so as to represent a characteristic density at a 
characteristic radius, $r=r_{\rm s}$.
The characteristic radius describes a typical scale at which the profile 
slope changes from the outer value to the inner value. 
Often it is close to the radius at which the slope is $\alpha=2$
(Navarro et al. 1996, 1997; Moore et al. 1998).

Since the dark matter density profile is self-similar, the dark matter 
mass profile is also self-similar. 
The dark matter mass enclosed within a radius $r$ is
%%%%%%%%%%%%%%%%%%%%%%%%%%%%%%%%%%%%%%%%%%%%%%%%%%%%%%%%%%%%%%%%%%%
\begin{equation}
 \label{eq:mass}
  M(\leq r)\equiv 4\pi\rho_{\rm s} r_{\rm s}^3 m\left(r/r_{\rm s}\right),
\end{equation}
%%%%%%%%%%%%%%%%%%%%%%%%%%%%%%%%%%%%%%%%%%%%%%%%%%%%%%%%%%%%%%%%%%%
where $m(x)$ is a non-dimensional mass profile given by
%%%%%%%%%%%%%%%%%%%%%%%%%%%%%%%%%%%%%%%%%%%%%%%%%%%%%%%%%%%%%%%%%%%
\begin{equation}
 \label{eq:m}
  m(x)\equiv \int_0^x du~u^2 y_{\rm dm}(u).
\end{equation}
%%%%%%%%%%%%%%%%%%%%%%%%%%%%%%%%%%%%%%%%%%%%%%%%%%%%%%%%%%%%%%%%%%% 
$y_{\rm dm}(u)$ must be shallower than $u^{-3}$ in the central region
in order for the mass to converge.

We define total dark matter mass, $M_{\rm vir}$, as the mass within 
the virial radius, $r_{\rm vir}$, i.e., $M_{\rm vir}\equiv M(\leq c)$,
where 
%%%%%%%%%%%%%%%%%%%%%%%%%%%%%%%%%%%%%%%%%%%%%%%%%%%%%%%%%%%%%%%%%%%
\begin{equation}
 \label{eq:c}
  c\equiv \frac{r_{\rm vir}}{r_{\rm s}}
\end{equation}
%%%%%%%%%%%%%%%%%%%%%%%%%%%%%%%%%%%%%%%%%%%%%%%%%%%%%%%%%%%%%%%%%%%
is a dimensionless parameter called the concentration parameter.
Evaluating equation (\ref{eq:mass}) at the virial radius,
we fix the normalization factor, $\rho_{\rm s}$, as
%%%%%%%%%%%%%%%%%%%%%%%%%%%%%%%%%%%%%%%%%%%%%%%%%%%%%%%%%%%%%%%%%%%
\begin{equation}
 \label{eq:rho0}
  \rho_{\rm s} = c^3\frac{M_{\rm vir}}{4\pi r_{\rm vir}^3 m(c)}.
\end{equation}
%%%%%%%%%%%%%%%%%%%%%%%%%%%%%%%%%%%%%%%%%%%%%%%%%%%%%%%%%%%%%%%%%%%

We calculate the virial radius, $r_{\rm vir}\left(M_{\rm vir},z\right)$,
with the spherical collapse model \cite{Peebles80}, 
%%%%%%%%%%%%%%%%%%%%%%%%%%%%%%%%%%%%%%%%%%%%%%%%%%%%%%%%%%%%%%%%%%%
\begin{equation}
 \label{eq:rvir}
  r_{\rm vir}
  = \left[
     \frac{M_{\rm vir}}
     {(4\pi/3)\Delta_{\rm c}(z)\rho_{\rm c}(z)}
   \right]^{1/3},
\end{equation}
%%%%%%%%%%%%%%%%%%%%%%%%%%%%%%%%%%%%%%%%%%%%%%%%%%%%%%%%%%%%%%%%%%%
where $\Delta_{\rm c}(z)$ is a spherical overdensity of
the virialized halo within $r_{\rm vir}$ at $z$, in units of 
the critical density of the universe at $z$, $\rho_{\rm c}(z)$.
In the Einstein--de Sitter universe model, $\Delta_{\rm c}(z)$
is independent of $z$, and is exactly $18\pi^2\simeq 178$.
Lacey \& Cole \shortcite{LC93} and Nakamura \& Suto \shortcite{NS97} 
provide $\Delta_{\rm c}(z)$ for arbitrary cosmological models.
In our cosmological model with $\Omega_{\rm m}=0.3$ and 
$\Omega_{\Lambda}=0.7$, the spherical overdensity of the virialized
halo at the present is 100: $\Delta_{\rm c}(z=0)=100$.

The only parameter in our framework is the concentration
parameter, $c$. 
Many $N$-body simulations show that $c$ decreases gradually 
with the virial mass.
Assuming the mass dependence of $c$ known, for example, 
from $N$-body simulations 
(Bullock et al. 2001; Eke, Navarro \& Steinmetz 2000), 
or from the non-linear dark matter power spectrum \cite{Seljak00}, 
one can predict $\rho_{\rm dm}(r)$ for any $M_{\rm vir}$, any $z$, 
and in principle any cosmological models.
This is the main advantage of the universal dark matter density profile
paradigm, and explains its popularity in recent work. 
Although the universal profile does not explain the dark matter 
profiles on the object-to-object basis \cite{JS00,KKBP01}, 
it is still successful in fitting the averaged form of the dark matter 
density profiles with reasonable accuracy.

So far, we have discussed the functional dependence of the 
profile on the characteristic radius. 
For a complete description, however, we need the functional form 
of $y_{\rm dm}(x)$. 
We adopt
%%%%%%%%%%%%%%%%%%%%%%%%%%%%%%%%%%%%%%%%%%%%%%%%%%%%%%%%%%%%%%%%%%%
\begin{equation}
 \label{eq:NFWprofiles}
  y_{\rm dm}(x) = \frac1{x^\alpha(1+x)^{3-\alpha}}.
\end{equation}
%%%%%%%%%%%%%%%%%%%%%%%%%%%%%%%%%%%%%%%%%%%%%%%%%%%%%%%%%%%%%%%%%%%
The asymptotic profile in the $x\gg 1$ regime is 
$y_{\rm dm}(x\gg 1) = x^{-3}$. 
This is the most common form found in 
the $N$-body simulations, although scatter around this is 
quite significant \cite{Thomas01}.
In the $x\ll 1$ regime, we have $y_{\rm dm}(x\ll 1) = x^{-\alpha}$.
Particularly, $\alpha=1$ corresponds to the profile proposed by 
Navarro et al. (1996, 1997), while $\alpha=3/2$ corresponds to the 
one proposed by Moore et al. \shortcite{Moore98} and 
Jing \& Suto \shortcite{JS00}. 
We will focus on these two particular cases throughout this paper.

Using these parameters, Suto et al. \shortcite{SSM98} have evaluated the 
relevant integrals for $\alpha=1$:
%%%%%%%%%%%%%%%%%%%%%%%%%%%%%%%%%%%%%%%%%%%%%%%%%%%%%%%%%%%%%%%%%%%
\begin{equation}
 \label{eq:m(x)NFW}
  m(x) = \ln(1+x)-\frac{x}{1+x},
\end{equation}
%%%%%%%%%%%%%%%%%%%%%%%%%%%%%%%%%%%%%%%%%%%%%%%%%%%%%%%%%%%%%%%%%%%
%%%%%%%%%%%%%%%%%%%%%%%%%%%%%%%%%%%%%%%%%%%%%%%%%%%%%%%%%%%%%%%%%%%
\begin{equation}
 \label{eq:f(x)NFW}
  \int_0^x du \frac{m(u)}{u^2} = 1-\frac{\ln(1+x)}x,
\end{equation}
%%%%%%%%%%%%%%%%%%%%%%%%%%%%%%%%%%%%%%%%%%%%%%%%%%%%%%%%%%%%%%%%%%%
and for $\alpha=3/2$:
%%%%%%%%%%%%%%%%%%%%%%%%%%%%%%%%%%%%%%%%%%%%%%%%%%%%%%%%%%%%%%%%%%%
\begin{equation}
 \label{eq:m(x)JS}
  m(x) = 2\ln\left(\sqrt{x}+\sqrt{1+x}\right)
        -2\sqrt{\frac{x}{1+x}},
\end{equation}
%%%%%%%%%%%%%%%%%%%%%%%%%%%%%%%%%%%%%%%%%%%%%%%%%%%%%%%%%%%%%%%%%%%
%%%%%%%%%%%%%%%%%%%%%%%%%%%%%%%%%%%%%%%%%%%%%%%%%%%%%%%%%%%%%%%%%%%
\begin{equation}
 \label{eq:f(x)JS}
  \int_0^x du \frac{m(u)}{u^2} = 
         -\frac{2\ln\left(\sqrt{x}+\sqrt{1+x}\right)}x
        +2\sqrt{\frac{1+x}{x}}.
\end{equation}
%%%%%%%%%%%%%%%%%%%%%%%%%%%%%%%%%%%%%%%%%%%%%%%%%%%%%%%%%%%%%%%%%%%
We use $\int_0^x du u^{-2}m(u)$ in next section.

There are several different empirical fitting formulae for the concentration
parameter in the literature. 
A recent compilation is Eke et al. \shortcite{ENS00}. 
For $\alpha=1$, we will use the one of Seljak \shortcite{Seljak00},
%%%%%%%%%%%%%%%%%%%%%%%%%%%%%%%%%%%%%%%%%%%%%%%%%%%%%%%%%%%%%%%%%%%
\begin{equation}
 \label{eq:concentration}
  c=6\left(\frac{M{\rm vir}}{10^{14}h^{-1}M_{\sun}}\right)^{-1/5}.
\end{equation}
%%%%%%%%%%%%%%%%%%%%%%%%%%%%%%%%%%%%%%%%%%%%%%%%%%%%%%%%%%%%%%%%%%%
For $\alpha=3/2$, we reduce the concentration parameter by a 
factor of 1.7 \cite{Seljak00,KKBP01}. 

\subsection{Overdensity radius and mass}

X-ray observations are not sufficiently sensitive 
to measure X-ray surface brightness and gas temperature profiles
out to the virial radius.
Therefore, instead of measuring the virial mass, many authors measure 
the mass within an overdensity radius, $r_\delta$, at which 
the dark matter density is $\delta$ times the critical density of the 
universe:
%%%%%%%%%%%%%%%%%%%%%%%%%%%%%%%%%%%%%%%%%%%%%%%%%%%%%%%%%%%%%%%%%%%
\begin{equation}
 \label{eq:rdelta}
  r_{\delta}
  \equiv \left[
     \frac{M(\leq r_\delta)}
     {(4\pi/3)\delta\rho_{\rm c}(z)}
   \right]^{1/3}.
\end{equation}
%%%%%%%%%%%%%%%%%%%%%%%%%%%%%%%%%%%%%%%%%%%%%%%%%%%%%%%%%%%%%%%%%%%
Note that the virial radius at $z=0$ corresponds to $r_{100}$ in our 
cosmological model, as $\Delta_{\rm c}(0)=100$ in equation (\ref{eq:rvir}).
Since we assume the dark matter density profile, and hence the mass 
profile, known, we can relate $M_{\rm vir}$ to $M(\leq r_\delta)$ by 
solving the equation
%%%%%%%%%%%%%%%%%%%%%%%%%%%%%%%%%%%%%%%%%%%%%%%%%%%%%%%%%%%%%%%%%%%
\begin{equation}
 \label{eq:delta2vir}
  M(\leq r_\delta)
  =
  \left[\frac{m(r_\delta/r_{\rm s})}{m(c)}\right]M_{\rm vir}
  =
  \left[\frac{m(cr_\delta/r_{\rm vir})}{m(c)}\right]M_{\rm vir}.
\end{equation}
%%%%%%%%%%%%%%%%%%%%%%%%%%%%%%%%%%%%%%%%%%%%%%%%%%%%%%%%%%%%%%%%%%%

\begin{figure}
  \begin{center}
    \leavevmode\epsfxsize=8.4cm \epsfbox{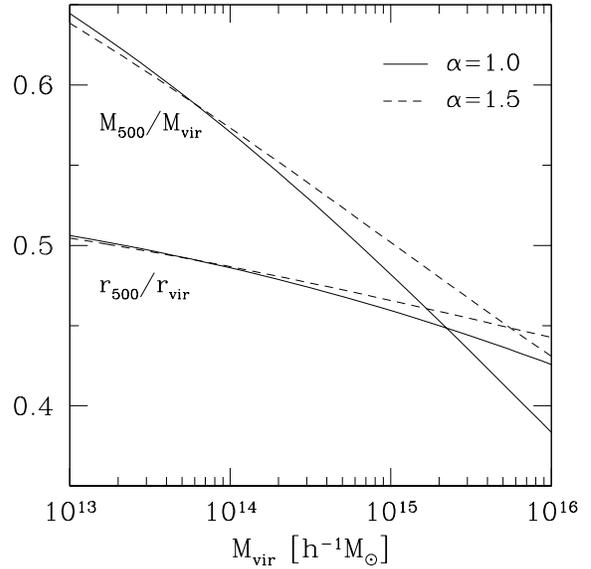}
  \end{center}
\caption{
 Ratio of the virial mass, $M_{\rm vir}$, to the overdensity mass at 
 $r_{500}$, $M_{500}\equiv M(\leq r_{500})$, as a function of $M_{\rm vir}$.
 Ratio of $r_{500}$ to the virial radius, $r_{\rm vir}$, is also shown.
 The solid lines represent $\alpha=1$, while the dashed lines 
 represent $\alpha=3/2$.}
\label{fig:M500_Mvir}
\end{figure}

Figure~\ref{fig:M500_Mvir} plots $M(\leq r_{500})$ as a function of 
$M_{\rm vir}$ for $\alpha=1$ and $3/2$. 
The mass dependence arises solely from the mass dependence of the 
concentration parameter given by equation (\ref{eq:concentration}). 
Since we will primarily be looking at haloes with the mass between 
$10^{13}h^{-1}M_{\sun}$ and $10^{15}h^{-1}M_{\sun}$, the relation 
between the two is roughly $M(\leq r_{500})\sim M_{\rm vir}/2$.
The mass measurement with the gas density profiles out to $r_{500}$ 
is thought to be accurate (e.g. Evrard, Metzler \& Navarro 1996);
this is why this radius has been frequently used in the literature.
Therefore, we will use $r_{500}$ and $M(\leq r_{500})$ for comparison 
of our theoretical predictions with observations.

Figure~\ref{fig:M500_Mvir} also plots a ratio of $r_{500}$ to the virial
radius as a function of $M_{\rm vir}$.
One finds that $r_{500}$ is half the virial radius. 
Note that there is some 
ambiguity in the definition of the virial radius.
Some authors use a fixed overdensity, $\delta$, regardless of cosmology, and
others use the spherical collapse value, $\Delta_{\rm c}$, that we use. 

Although we will continue to use the concentration parameter defined by
equation~(\ref{eq:c}), we may also introduce
a more model-independent quantity like $c_{500}=r_{500}/r_{\rm s}$, 
which does not depend on the definition of the virial radius. 
For the cosmology used here, $c_{500}$ is half the concentration
parameter. 

%%%%%%%%%%%%%%%%%%%%%%%%%%%%%%%%%%%%%%%%%%%%%%%%%%%%%%%%%%%%%%%%%%%%%%
%%%%%%%%%%%%%%%%%%%%%%%%%%%%%%%%%%%%%%%%%%%%%%%%%%%%%%%%%%%%%%%%%%%%%%

%%%%%%%%%%%%%%%%%%%%%%%%%%%%%%%%%%%%%%%%%%%%%%%%%%%%%%%%%%%%%%%%%%%
\section{Self-similar Gas Density and Temperature Profiles}

\subsection{Polytropic gas model in hydrostatic equilibrium}

Since we assume the dark matter density profile self-similar, 
a gas density profile, $\rho_{\rm gas}(r)$, would also be self-similar, 
if no additional scale is introduced 
(this translates into a requirement that the density-temperature 
relation is a power law). 
Hence, we have
%%%%%%%%%%%%%%%%%%%%%%%%%%%%%%%%%%%%%%%%%%%%%%%%%%%%%%%%%%%%%%%%%%%
\begin{equation}
 \label{eq:gasprofile}
  \rho_{\rm gas}(r)= \rho_{\rm gas}(0) y_{\rm gas}(r/r_{\rm s}),
\end{equation}
%%%%%%%%%%%%%%%%%%%%%%%%%%%%%%%%%%%%%%%%%%%%%%%%%%%%%%%%%%%%%%%%%%%
where $\rho_{\rm gas}(0)$ is the gas density at $r=0$.
We parameterize the gas density profile in this way because, 
in contrast to the dark matter density profile, 
a flat gas density core near the centre of haloes
is well observed, and is expected if the gas temperature is not a 
strong function of radii. 
Makino et al. \shortcite{MSS98} and Suto et al. \shortcite{SSM98} show 
that the hydrostatic equilibrium between the gas pressure and 
the self-similar dark matter potential gives 
the self-similar density profile. 

To take phenomenologically into account the effect of the gas 
temperature gradient which is found both in observations and in 
hydrodynamic simulations, Suto et al. \shortcite{SSM98} parameterize 
the gas pressure in a polytropic form, 
$P_{\rm gas}\propto \rho_{\rm gas}T_{\rm gas}
            \propto \rho^\gamma_{\rm gas}$.
This parameterization gives a self-similar temperature profile,
%%%%%%%%%%%%%%%%%%%%%%%%%%%%%%%%%%%%%%%%%%%%%%%%%%%%%%%%%%%%%%%%%%%
\begin{equation}
 \label{eq:temprof} 
  T_{\rm gas}(r/r_{\rm s})= T_{\rm gas}(0)y_{\rm gas}^{\gamma-1}(r/r_{\rm s}).
\end{equation}
%%%%%%%%%%%%%%%%%%%%%%%%%%%%%%%%%%%%%%%%%%%%%%%%%%%%%%%%%%%%%%%%%%%
Then, the hydrostatic equilibrium equation,
%%%%%%%%%%%%%%%%%%%%%%%%%%%%%%%%%%%%%%%%%%%%%%%%%%%%%%%%%%%%%%%%%%%
\begin{equation}
  \rho_{\rm gas}^{-1}{dP_{\rm gas} \over dr} = -G {M(\leq r) \over r^2}, 
\end{equation}
%%%%%%%%%%%%%%%%%%%%%%%%%%%%%%%%%%%%%%%%%%%%%%%%%%%%%%%%%%%%%%%%%%%
gives the differential equation for the gas profile, $y_{\rm gas}(x)$, 
%%%%%%%%%%%%%%%%%%%%%%%%%%%%%%%%%%%%%%%%%%%%%%%%%%%%%%%%%%%%%%%%%%%
\begin{equation}
 \label{eq:hydro}
  \frac{d y_{\rm gas}^{\gamma-1}(r/r_{\rm s})}{dr}
  =
  -\left(\frac{\gamma-1}{\gamma}\right)
  \frac{G\mu m_{\rm p}M_{\rm vir}}{k_{\rm B}T_{\rm gas}(0) r^2}\left[\frac{m(r/r_{\rm s})}{m(c)}\right].
\end{equation}
%%%%%%%%%%%%%%%%%%%%%%%%%%%%%%%%%%%%%%%%%%%%%%%%%%%%%%%%%%%%%%%%%%%
Here, $M_{\rm vir}$ is the virial mass, and $m(x)$ is the dimensionless
mass profile given by equation (\ref{eq:m}).
Since $y_{\rm gas}(0)=1$ (see equation~(\ref{eq:gasprofile})), 
equation (\ref{eq:hydro}) is formally solved for $y_{\rm gas}(x)$ 
as \cite{SSM98} 
%%%%%%%%%%%%%%%%%%%%%%%%%%%%%%%%%%%%%%%%%%%%%%%%%%%%%%%%%%%%%%%%%%%
\begin{equation}
 \label{eq:solution}
  y^{\gamma-1}_{\rm gas}(x)
  = 1-3\eta^{-1}(0)\left(\frac{\gamma-1}{\gamma}\right)
  \left[\frac{c}{m(c)}\right]
     \int_0^{x} du\frac{m(u)}{u^2},
\end{equation}
%%%%%%%%%%%%%%%%%%%%%%%%%%%%%%%%%%%%%%%%%%%%%%%%%%%%%%%%%%%%%%%%%%%
where 
%%%%%%%%%%%%%%%%%%%%%%%%%%%%%%%%%%%%%%%%%%%%%%%%%%%%%%%%%%%%%%%%%%%
\begin{equation}
 \label{eq:eta}
  \eta^{-1}(x)\equiv 
  \frac{G\mu m_{\rm p}M_{\rm vir}}{3r_{\rm vir}k_{\rm B}T_{\rm gas}(x)}
\end{equation}
%%%%%%%%%%%%%%%%%%%%%%%%%%%%%%%%%%%%%%%%%%%%%%%%%%%%%%%%%%%%%%%%%%%
is the normalization factor relating $T_{\rm gas}(x)$ to $M_{\rm vir}$.
Note that $\eta(x)= \eta(0)y_{\rm gas}^{\gamma-1}(x)$.

The hydrostatic equation (\ref{eq:hydro}) does not include 
$\rho_{\rm gas}(0)$, but $\eta(0)$ representing $T_{\rm gas}(0)$; 
thus, $\eta(0)$ is one of the two free parameters in the model. 
The other parameter is the polytropic index, $\gamma$. 
As we will show below, by requiring the gas and the dark matter profiles 
to agree outside the core, we can determine both the parameters.

%%%%%%%%%%%%%%%%%%%%%%%%%%%%%%%%%%%%%%%%%%%%%%%%%%%%%%%%%%%%%%%%%%%
\subsection{Gas tracing dark matter outside the core}

We assume that the gas density profile traces the dark matter density 
profile in the outer region of the halo. 
This assumption is the main ingredient of our model presented in this paper.

Qualitatively, this assumption is reasonable, as the gas and the dark 
matter obey similar equations of motion to each other. 
The main difference is that the gas falling into the cluster
potential well is shocked and then heated, while the dark matter, being 
collisionless, does not undergo shock heating. 
What it does undergo is orbit mixing, which causes multiple streams to 
appear at the same physical position. 
This gives the velocity dispersion, an effective
temperature of the dark matter. 
Taking moments of the collisionless Boltzmann equation leads to the 
Jeans equations \cite{BT87} which are formally similar to the gas fluid 
equations. 
The main difference is that, for the collisionless system, the 
effective pressure is the velocity dispersion tensor that can be 
anisotropic, while the gas pressure is isotropic.

An even better reason for our assumption is that many hydrodynamic simulations 
have observed the gas density profile tracing the dark matter density profile
outside the core, $r\ga r_{\rm vir}/2$
(Navarro, Frenk \& White 1995; Bryan \& Norman 1998; Eke, 
Navarro \& Frenk 1998; Frenk et al. 1999; 
Pearce et al. 2000; Lewis et al. 2000; Yoshikawa, Jing \& Suto 2000).
For example, the Santa Barbara cluster project \cite{Santa99}, 
the comparison between a dozen numerical codes for a single cluster, 
shows this property with 10\% or better accuracy.
Other larger samples also show the agreement. 
While additional physics missing in the non-radiative simulations, 
such as cooling and star formation, may change the structure of the gas,
the outer regions appear to remain relatively unaffected
\cite{PTCE00,Lewis00}.

We require the gas density profile, $y_{\rm gas}(x)$, to match the 
dark matter density profile, $y_{\rm dm}(x)$, at a certain matching 
point, $x_*$.  
We impose this by requiring slopes of these two profiles to be the same
at $x_*$,
%%%%%%%%%%%%%%%%%%%%%%%%%%%%%%%%%%%%%%%%%%%%%%%%%%%%%%%%%%%%%%%%%%%
\begin{equation}
 \label{eq:condition}
  s_*\equiv
  \left.\frac{d\ln y_{\rm dm}(x)}{\ln x}\right|_{x=x_*}
  =
  \left.\frac{d\ln y_{\rm gas}(x)}{\ln x}\right|_{x=x_*},
\end{equation}
%%%%%%%%%%%%%%%%%%%%%%%%%%%%%%%%%%%%%%%%%%%%%%%%%%%%%%%%%%%%%%%%%%%
where $s_*$ denotes an effective slope of the dark matter density
profile at $x_*$.
Substituting $y_{\rm dm}(x)$ for equation~(\ref{eq:NFWprofiles}), 
we obtain the effective slope at $x_*$,
%%%%%%%%%%%%%%%%%%%%%%%%%%%%%%%%%%%%%%%%%%%%%%%%%%%%%%%%%%%%%%%%%%%
\begin{equation}
 \label{eq:sstar}
  s_* = -\left[\alpha+(3-\alpha)\frac{x_*}{1+x_*}\right].
\end{equation}
%%%%%%%%%%%%%%%%%%%%%%%%%%%%%%%%%%%%%%%%%%%%%%%%%%%%%%%%%%%%%%%%%%%
This constraint, equation~(\ref{eq:condition}), fixes one of the two 
parameters of the model, $\eta(0)$, the normalization factor
of the mass--temperature relation given by equation (\ref{eq:eta}).

Substituting $y_{\rm gas}(x)$ in  equation~(\ref{eq:condition})
for equation (\ref{eq:solution}), we obtain a solution for 
the mass--temperature normalization factor at the centre, $\eta(0)$, as
%%%%%%%%%%%%%%%%%%%%%%%%%%%%%%%%%%%%%%%%%%%%%%%%%%%%%%%%%%%%%%%%%%%
\begin{eqnarray}
 \nonumber
  \eta(0)
  &=&
  \gamma^{-1}
  \left\{
  \left(\frac{-3}{s_*}\right)
  \left[ \frac{x_*^{-1}m(x_*)}{c^{-1}m(c)} \right]\right.\\
 \label{eq:eta0}
  & &\qquad \left. +3(\gamma-1)\left[\frac{c}{m(c)}\right]
  \int_0^{x_*} du \frac{m(u)}{u^2}
 \right\}.
\end{eqnarray}
%%%%%%%%%%%%%%%%%%%%%%%%%%%%%%%%%%%%%%%%%%%%%%%%%%%%%%%%%%%%%%%%%%%
In the isothermal limit, $\gamma\rightarrow 1$, we have
%%%%%%%%%%%%%%%%%%%%%%%%%%%%%%%%%%%%%%%%%%%%%%%%%%%%%%%%%%%%%%%%%%%
\begin{equation}
 \label{eq:eta_iso}
  \eta(0)=
  \left(\frac{-3}{s_*}\right)
 \left[ \frac{x_*^{-1}m(x_*)}{c^{-1}m(c)} \right].
\end{equation}
%%%%%%%%%%%%%%%%%%%%%%%%%%%%%%%%%%%%%%%%%%%%%%%%%%%%%%%%%%%%%%%%%%%
Note that this gives $\eta(0)=1$ for $x_*=c$ in the limit of large 
concentration, $c\rightarrow \infty$.

In order for the two profiles to agree over a wide range, 
the two slopes should agree not only at $x=x_*$, but everywhere in 
the outer region of haloes, so within a factor of two above and 
below the virial radius ($c/2<x<2c$) according to the simulations. 
In this way, the solution for $\eta(0)$ will not depend on $x_*$. 
This requirement will not be satisfied in general; however, we still 
have one free parameter left, the polytropic index, $\gamma$. 
We fix it by requiring the solution for $\eta(0)$ not to depend on $x_*$. 
Our model then no longer contains any free parameters.
Using the solution for $\eta(0)$, we calculate the gas temperature 
at an arbitrary radial point with $\eta(r)=\eta(0)y^{\gamma-1}_{\rm gas}(r)$.

\begin{figure}
  \begin{center}
    \leavevmode\epsfxsize=8.4cm \epsfbox{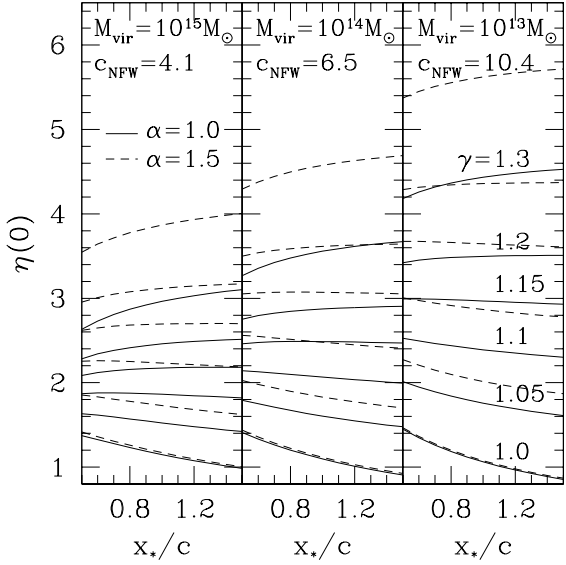}
  \end{center}
\caption{
 Mass--temperature normalization factor at the centre, $\eta(0)$, 
 predicted by the polytropic gas model with a polytropic index of 
 $\gamma$. 
 $\eta(0)$ is plotted as a function of the matching point, $x_*$,
 in units of the concentration parameter, $c$. 
 The solid lines represent $\alpha=1$, 
 while the dashed lines represent $\alpha=3/2$. 
 From left to right panels, $M_{\rm vir}=10^{15}$, 
 $10^{14}$, and $10^{13}~M_{\sun}$ are plotted.
 Each mass corresponds to the concentration parameter of $c_{\rm NFW}=4.1$, 
 6.5, and 10.4 for $\alpha=1$, respectively, 
 while $c=c_{\rm NFW}/1.7$ for $\alpha=3/2$.
 In each panel, $\gamma=1.0$, 1.05, 1.1, 1.15, 1.2 and 1.3 are 
 plotted from bottom to top lines.}
\label{fig:eta0_xstar}
\end{figure}

Figure~\ref{fig:eta0_xstar} plots $\eta(0)$ as a function of the matching 
point, $x_*$, for a variety of polytropic indices, $\gamma$.
It follows from this figure that $\gamma=1.1-1.2$ satisfies the 
requirement that $\eta(0)$ is independent of $x_*$. 
The isothermal case, $\gamma=1$, fails this requirement; 
the isothermal gas density profile cannot be made similar to the 
dark matter density profile in the outer parts of the halo.  

The best-fitting value of $\gamma$ depends on $c$ very weakly, changing from 
$1.1$ to $1.2$ as $c$ varies from 4 to 10 (or the mass from 
$M_{\rm vir}=10^{15}M_{\sun}$ to $10^{13}M_{\sun}$).
We find a simple linear fit for $\gamma=\gamma(c)$, 
%%%%%%%%%%%%%%%%%%%%%%%%%%%%%%%%%%%%%%%%%%%%%%%%%%%%%%%%%%%%%%%%%%%
\begin{equation}
 \label{eq:bestgamma}
  \gamma=1.15 + 0.01\left(c_{\rm NFW}-6.5\right),
\end{equation}
%%%%%%%%%%%%%%%%%%%%%%%%%%%%%%%%%%%%%%%%%%%%%%%%%%%%%%%%%%%%%%%%%%%
where $c_{\rm NFW}$ denotes the concentration parameter for 
the dark matter density profile with $\alpha=1$.
Equation (\ref{eq:bestgamma}) gives an adequate value to both 
$\alpha=1$ and $\alpha=3/2$ profiles.
Hence, $\gamma$ is no longer free, but fixed.
Here, note that the concentration for $\alpha=3/2$ is given by 
$c=c_{\rm NFW}/1.7$.

%%%%%%%%%%%%%%%%%%%%%%%%%%%%%%%%%%%%%%%%%%%%%%%%%%%%%%%%%%%%%%%%%%%%%%
%%%%%%%%%%%%%%%%%%%%%%%%%%%%%%%%%%%%%%%%%%%%%%%%%%%%%%%%%%%%%%%%%%%%%%

\begin{figure}
  \begin{center}
    \leavevmode\epsfxsize=8.4cm \epsfbox{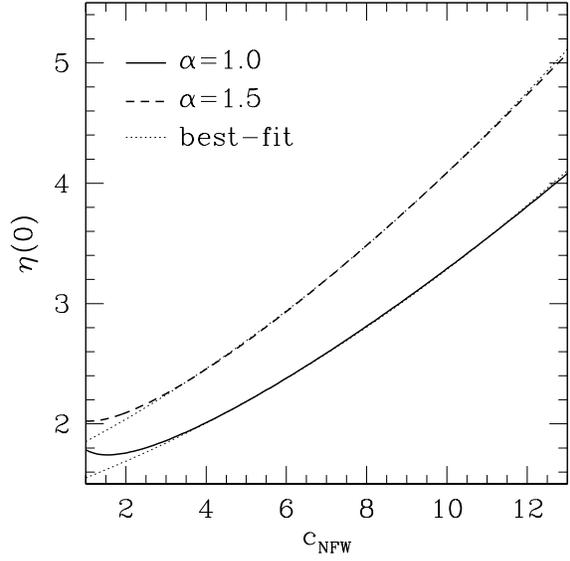}
  \end{center}
\caption{
 Mass--temperature normalization factor at the centre, $\eta(0)$,
 as a function of the concentration parameter for $\alpha=1$, $c_{\rm NFW}$. 
 The solid line represents $\alpha=1$, while the dashed line 
 represents $\alpha=3/2$.
 The dotted lines are the best-fitting lines given by equations
 (\ref{eq:fit1}) and (\ref{eq:fit3/2}).}
\label{fig:eta0_c}
\end{figure}

Furthermore, combining equation (\ref{eq:eta0}) with (\ref{eq:bestgamma})
gives $\eta(0)$ as a function of the concentration parameter.  
Figure~\ref{fig:eta0_c} plots $\eta(0)$ as a function of $c_{\rm NFW}$.
We find that the dependence of $\eta(0)$ on $c_{\rm NFW}$ is well fitted by
%%%%%%%%%%%%%%%%%%%%%%%%%%%%%%%%%%%%%%%%%%%%%%%%%%%%%%%%%%%%%%%%%%%
\begin{equation}
 \label{eq:fit1}
  \eta(0)= 0.00676\left(c_{\rm NFW}-6.5\right)^2
          + 0.206 \left(c_{\rm NFW}-6.5\right) + 2.48,
\end{equation}
%%%%%%%%%%%%%%%%%%%%%%%%%%%%%%%%%%%%%%%%%%%%%%%%%%%%%%%%%%%%%%%%%%%
for $\alpha=1$, and
%%%%%%%%%%%%%%%%%%%%%%%%%%%%%%%%%%%%%%%%%%%%%%%%%%%%%%%%%%%%%%%%%%%
\begin{equation}
 \label{eq:fit3/2}
  \eta(0)= 0.00776\left(c_{\rm NFW}-6.5\right)^2
         + 0.264 \left(c_{\rm NFW}-6.5\right) + 3.07,
\end{equation}
%%%%%%%%%%%%%%%%%%%%%%%%%%%%%%%%%%%%%%%%%%%%%%%%%%%%%%%%%%%%%%%%%%%
for $\alpha=3/2$; thus, the higher concentration implies the central gas
temperature being higher than the gas temperature at the virial radius.
Since by construction the particular choice of $x_*$ does not 
affect our analysis, we take $x_*=c$ (or $r_*=r_{\rm vir}$) for definiteness 
in the following.

%%%%%%%%%%%%%%%%%%%%%%%%%%%%%%%%%%%%%%%%%%%%%%%%%%%%%%%%%%%%%%%%%%%%%%
%%%%%%%%%%%%%%%%%%%%%%%%%%%%%%%%%%%%%%%%%%%%%%%%%%%%%%%%%%%%%%%%%%%%%%

\begin{figure}
  \begin{center}
    \leavevmode\epsfxsize=8.4cm \epsfbox{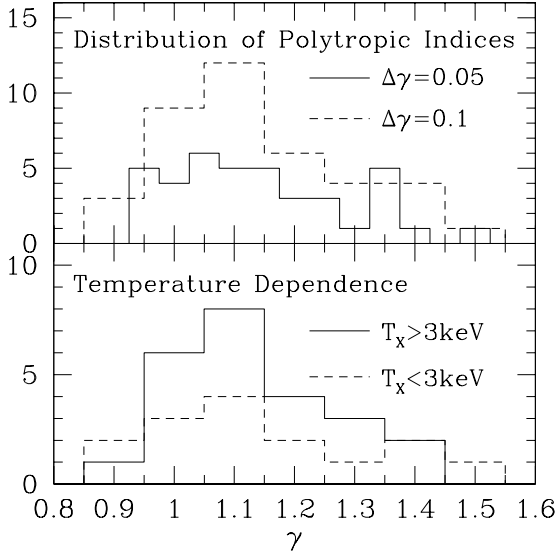}
  \end{center}
\caption{
 Measured distribution of the polytropic index, $\gamma$,
 from Finoguenov et al. \shortcite{FRB01}.
 The top panel plots the whole sample in two different bins:
 the solid line is $\Delta\gamma=0.1$ bin, and the dashed 
 line is $\Delta\gamma=0.05$ bin.
 The mean value of $\gamma$ is 1.15.
 The bottom panel plots the temperature dependence of $\gamma$.
 The solid line shows the clusters above 3~keV, 
 while the dashed line shows the cooler ones below 3~keV.
 There is no clear evidence that the polytropic index changes with 
 temperature.}
\label{fig:gamma}
\end{figure}

Figure~\ref{fig:gamma} shows histograms of observationally measured 
polytropic indices. 
The top panel plots the whole sample from Finoguenov et al. \shortcite{FRB01}.
There is a peak in the population around $\gamma\simeq 1.1$.
The mean value is $\gamma=1.15$.
This mean value is close to our best-fitting value given by equation 
(\ref{eq:bestgamma}).
The bottom panel of this figure plots the hotter objects than 3~{\rm keV} in
the solid line, and the cooler ones below it in the dashed line.
The distribution is basically the same for these different temperatures;
thus, there is no significant dependence of the polytropic index
on the gas temperature. 
This also agrees with our model.

%%%%%%%%%%%%%%%%%%%%%%%%%%%%%%%%%%%%%%%%%%%%%%%%%%%%%%%%%%%%%%%%%%%%%%
%%%%%%%%%%%%%%%%%%%%%%%%%%%%%%%%%%%%%%%%%%%%%%%%%%%%%%%%%%%%%%%%%%%%%%

\subsection{Mean gas temperature of haloes}

Since X-ray observations do not measure the gas temperature itself
but the emission-weighted mean temperature, we use the latter 
as a representative gas temperature of the halo.
If the halo is isothermal, then the gas temperature and the 
emission-weighted mean temperature are the same; however,
the temperature gradient makes them different.
We predict the emission-weighted mean temperature, $T_{\rm X}$, with 
this spherical polytropic gas model as \cite{SSM98}
%%%%%%%%%%%%%%%%%%%%%%%%%%%%%%%%%%%%%%%%%%%%%%%%%%%%%%%%%%%%%%%%%%%
\begin{eqnarray}
  \nonumber
   T_{\rm X}
   &\equiv& 
   \frac{\int dV \rho^2_{\rm gas}\Lambda(T_{\rm gas})T_{\rm gas}}
   {\int dV \rho^2_{\rm gas}\Lambda(T_{\rm gas})}\\
  \label{eq:Tx}
   &\approx&
   T_{\rm gas}(0)\left[
		  \frac{\int_0^{x_{\rm max}} x^2 dx 
		  \left[y_{\rm gas}(x)\right]^{(3\gamma+1)/2}}
		  {\int_0^{x_{\rm max}} x^2 dx 
		  \left[y_{\rm gas}(x)\right]^{(\gamma+3)/2}}
		\right],
\end{eqnarray}
%%%%%%%%%%%%%%%%%%%%%%%%%%%%%%%%%%%%%%%%%%%%%%%%%%%%%%%%%%%%%%%%%%%
where we have approximated the cooling function, $\Lambda(T_{\rm gas})$, with  
the bolometric bremsstrahlung, 
$\Lambda(T_{\rm gas})\propto T_{\rm gas}^{1/2}\propto 
y_{\rm gas}^{(\gamma-1)/2}$.
Note that $\rho_{\rm gas}\propto y_{\rm gas}$, and 
$T_{\rm gas}\propto y_{\rm gas}^{\gamma-1}$.
Since we do not attempt to predict the X-ray luminosity but the X-ray 
temperature, this approximation for the cooling function seems sufficient.

\begin{figure}
  \begin{center}
    \leavevmode\epsfxsize=8.4cm \epsfbox{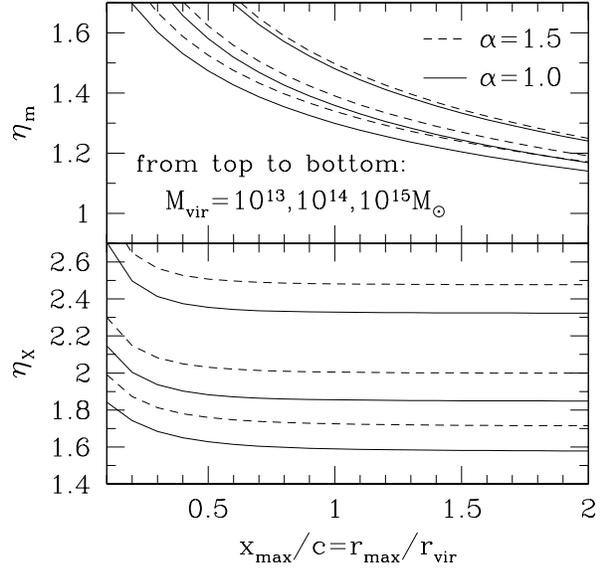}
  \end{center}
\caption{
 Predicted mass--temperature normalization factors 
 as a function of the upper boundary of the integral, $x_{\rm max}$, 
 in equations (\ref{eq:eta_x}) and (\ref{eq:eta_m}).
 The bottom panel is the emission-weighted normalization factor, while 
 the top panel is the mass-weighted one.
 The solid lines use the dark matter density profile for $\alpha=1$,
 while the dashed lines use the one for $\alpha=3/2$. 
 From top to bottom lines,  
 $M=10^{15}$, $10^{14}$, and $10^{13}M_{\sun}$ are shown.}
\label{fig:eta_xmax}
\end{figure}

For a practical evaluation of $T_{\rm X}$, we should care how to choose 
the integration boundary, $x_{\rm max}$.
As $y_{\rm gas}(x)\propto x^{s_*} \approx x^{-3}$ for $x\gg 1$, 
$x^3 \left[y_{\rm gas}(x)\right]^{(\gamma+3)/2}\propto x^{-3(\gamma+1)/2}
\approx x^{-3}$;
thus, the integral in the denominator of equation (\ref{eq:Tx}) 
converges rapidly as $x^{-3}$. 
The numerator converges even more rapidly than the denominator as long as 
$\gamma>1$.
This rapid convergence implies that a specific choice 
of $x_{\rm max}$ has no systematic effect on $T_{\rm X}$. 
Therefore, we can robustly compare $T_{\rm X}$ calculated from our model 
with observations as well as simulations.
The bottom panel of figure~\ref{fig:eta_xmax} plots the 
sensitivity of $T_{\rm X}$ to $x_{\rm max}$.
The plotted quantity, $\eta_{\rm X}$, is the emission-weighted 
mass--temperature normalization factor defined by equation 
(\ref{eq:eta_x}) below, and is equivalent to $T_{\rm X}$.
We find that $T_{\rm X}$ is quite insensitive to $x_{\max}$ for 
$x_{\rm max}\ga 0.2 c\sim 0.4 r_{500}/r_{\rm s}$. 
We use $x_{\rm max}=c$ for definiteness.

%%%%%%%%%%%%%%%%%%%%%%%%%%%%%%%%%%%%%%%%%%%%%%%%%%%%%%%%%%%%%%%%%%%%%%
%%%%%%%%%%%%%%%%%%%%%%%%%%%%%%%%%%%%%%%%%%%%%%%%%%%%%%%%%%%%%%%%%%%%%%

To compare our predictions with observations and simulations, 
we replace the normalization factor of the mass--temperature relation, 
$\eta(r)$ (equation~(\ref{eq:eta})), with the emission-weighted mean 
normalization factor,
%%%%%%%%%%%%%%%%%%%%%%%%%%%%%%%%%%%%%%%%%%%%%%%%%%%%%%%%%%%%%%%%%%%
\begin{eqnarray}
 \nonumber
  \eta_{\rm X}
  &\equiv& 
  \frac{\int dV \rho^2_{\rm gas}\Lambda(T_{\rm gas})\eta(r)}
  {\int dV \rho^2_{\rm gas}\Lambda(T_{\rm gas})}\\
 \label{eq:eta_x}
  &\approx&
  \eta(0)
  \left[
  \frac{\int_0^{x_{\rm max}} x^2 dx 
  \left[y_{\rm gas}(x)\right]^{(3\gamma+1)/2}}
  {\int_0^{x_{\rm max}} x^2 dx 
  \left[y_{\rm gas}(x)\right]^{(\gamma+3)/2}}\right].
\end{eqnarray}
%%%%%%%%%%%%%%%%%%%%%%%%%%%%%%%%%%%%%%%%%%%%%%%%%%%%%%%%%%%%%%%%%%%
Several hydrodynamic simulations have calculated this normalization 
factor \cite{EMN96,BN98,ENF98,YJS00,Thomas01}.
For example, $\eta_{\rm X}$ corresponds to 
the emission-weighted $\gamma$ in Yoshikawa et al. \shortcite{YJS00}, 
$(3/2)f_{\rm T}$ in Bryan \& Norman \shortcite{BN98}, 
and $(3/2)\overline{\beta}^{-1}_{\rm TM}$ in Eke et al. \shortcite{ENF98}.

Since the emission-weighted mean temperature is significantly weighted
toward the central region of haloes, it is sensitive
to the numerical resolution \cite{Lewis00}.
We find that there is significant dispersion of $\eta_{\rm X}$ among
different hydrodynamic simulations, in contrast to the mass-weighted 
mean normalization factor, $\eta_{\rm m}$, defined as \cite{YJS00,Thomas01}
%%%%%%%%%%%%%%%%%%%%%%%%%%%%%%%%%%%%%%%%%%%%%%%%%%%%%%%%%%%%%%%%%%%
\begin{equation}
 \label{eq:eta_m}
  \eta_{\rm m}
  \equiv 
  \frac{\int dV \rho_{\rm gas}\eta(r)}
  {\int dV \rho_{\rm gas}}
  =
  \eta(0)
  \left[
  \frac{\int_0^{x_{\rm max}} x^2 dx~ y_{\rm gas}^\gamma(x)}
  {\int_0^{x_{\rm max}} x^2 dx~ y_{\rm gas}(x)}\right].
\end{equation}
%%%%%%%%%%%%%%%%%%%%%%%%%%%%%%%%%%%%%%%%%%%%%%%%%%%%%%%%%%%%%%%%%%%
This is less weighted towards the central region of haloes than 
the emission-weighted normalization.
Unfortunately, $\eta_{\rm m}$ is not directly observed from 
the X-ray observations, while it is less sensitive to the numerical 
resolution of simulations. 
In contrast to the emission-weighted normalization, the denominator of 
equation (\ref{eq:eta_m}) does not converge, but diverges logarithmically. 
The numerator converges if $\gamma>1$, but very slowly. 
The top panel of figure~\ref{fig:eta_xmax} plots the sensitivity of 
$\eta_{\rm m}$ to $x_{\rm max}$. 
We find that $\eta_{\rm m}$ is rather sensitive to $x_{\rm max}$; 
thus, we use the virial radius for the integration
boundary: $x_{\rm max}=c$, to compare the predicted $\eta_{\rm m}$ with
simulations, for Thomas et al. \shortcite{Thomas01} have calculated
$\eta_{\rm m}$ within the virial radius.

\begin{figure}
  \begin{center}
    \leavevmode\epsfxsize=8.4cm \epsfbox{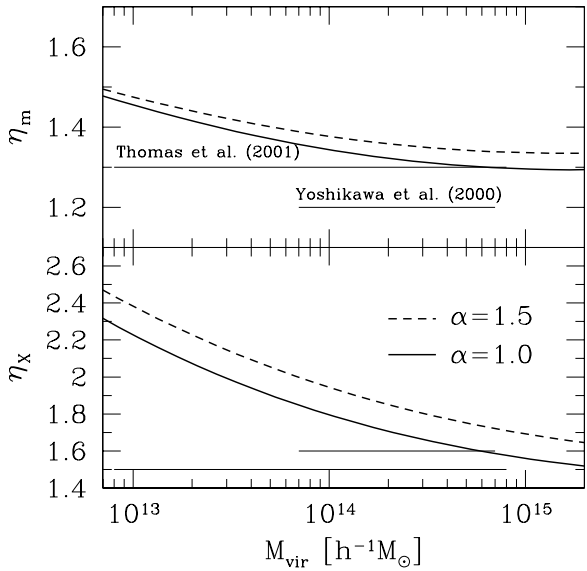}
  \end{center}
\caption{
 Predicted mass--temperature normalization factors 
 as a function of the virial mass, $M_{\rm vir}$.
 The emission-weighted normalization factor, $\eta_{\rm X}$ 
 (equation~(\ref{eq:eta_x})), is plotted in the bottom panel,
 while the mass-weighted one, $\eta_{\rm m}$ (equation~(\ref{eq:eta_m})), 
 is plotted in the top panel.
 The thick solid lines represent $\alpha=1$, while the thick dashed lines 
 represent $\alpha=3/2$.
 The thin solid lines indicate the normalization factors derived 
 from hydrodynamic simulations carried out by Thomas et al. 
 \shortcite{Thomas01} (upper line in the top panel and lower line in
 the bottom panel) and Yoshikawa et al. \shortcite{YJS00} 
 (lower line in the top panel and upper line in the bottom panel).}
\label{fig:eta_mass}
\end{figure}

Using equation~(\ref{eq:eta_x}) and (\ref{eq:eta_m}), we 
calculate $\eta_{\rm X}$ and $\eta_{\rm m}$ as a function of the halo mass. 
Figure~\ref{fig:eta_mass} compares the predictions with the simulations
for the dark matter profiles with $\alpha=1$ and 3/2.
To facilitate the comparison, we plot the dependence against the 
virial mass rather than the concentration parameter, but the two are 
related to each other through equation~(\ref{eq:concentration}).
We find that $\eta_{\rm X} > \eta_{\rm m}$, i.e., $T_{\rm X}>T_{\rm m}$, 
as observed in hydrodynamic simulations carried out by
Yoshikawa et al. \shortcite{YJS00} and Thomas et al. \shortcite{Thomas01}.
This is because the emission-weighted mean temperature is much more
weighted towards the central region than the mass-weighted one.
The gas temperature in the central region is higher than in the outer
region, as long as $\gamma>1$.

We also find that the normalization factors increase as the mass
decreases, more so for $\eta_{\rm X}$ than for $\eta_{\rm m}$. 
This is caused by the concentration dependence of the halo mass. 
The gas density in the core for higher $c$ (smaller mass) is larger than 
that for lower $c$ (larger mass). 
This in turn gives a smaller mass halo 
a higher temperature because of the assumed 
polytropic relation between the gas density and temperature. 

Many hydrodynamic simulations in the literature give
the mass--temperature normalization factors weighted by either the 
emission or the mass.
Figure~\ref{fig:eta_mass} compares our predictions with those obtained 
from the simulations.
Overall, the predicted normalizations agree with the 
simulations in the high mass regime (except for simulations by 
Bryan \& Norman \shortcite{BN98} who predict 20\% lower normalizations 
compared with the other simulations and with our results).

Our predicted emission-weighted normalizations are higher than the simulated
ones for smaller mass haloes. 
It is possible that a poorer numerical resolution for smaller mass haloes
causes the disagreement.
It is also possible that the mass dependence has not been noticed, 
as most of previous work has relied on a self-similar model for comparison, 
in which there is no mass dependence of the concentration parameter. 
The predicted mass-weighted normalizations which are less sensitive to the 
resolution issues agree better with the simulations.
We present the comparison with observations in the next section.
 
%%%%%%%%%%%%%%%%%%%%%%%%%%%%%%%%%%%%%%%%%%%%%%%%%%%%%%%%%%%%%%%%%%%%%%
%%%%%%%%%%%%%%%%%%%%%%%%%%%%%%%%%%%%%%%%%%%%%%%%%%%%%%%%%%%%%%%%%%%%%%

\subsection{Predicted gas density and gas temperature profiles}

\begin{figure}
  \begin{center}
    \leavevmode\epsfxsize=8.4cm \epsfbox{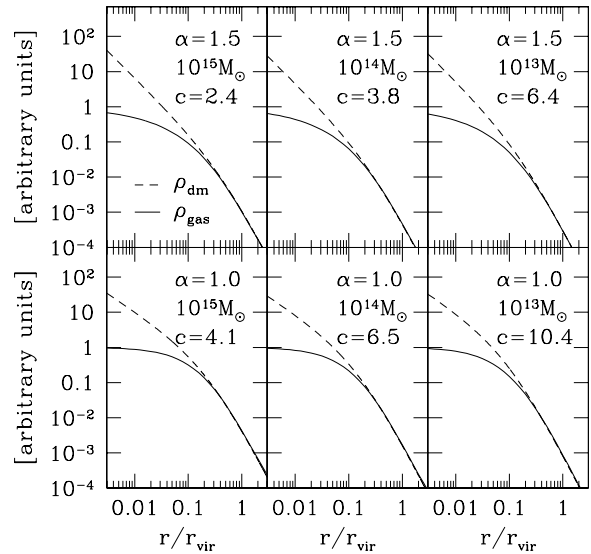}
  \end{center}
\caption{
 Predicted gas density profiles (solid lines) shown with universal 
 dark matter density profiles (dashed lines). 
 The top panels use the dark matter density profile for $\alpha=3/2$,
 while the bottom panels use the one for $\alpha=1$. 
 From left to right panels, each panel shows the case of 
 $M=10^{15}$, $10^{14}$, and $10^{13}M_{\sun}$, respectively. 
 The corresponding concentration parameter, $c$, to the mass is quoted.}
\label{fig:gasprofiles}
\end{figure}

We plot the dark matter (dashed lines) and the gas density (solid lines)
profiles in figure~\ref{fig:gasprofiles} for $M=10^{15}$, $10^{14}$, and 
$10^{13}M_{\sun}$ from left to right panels.
$\alpha=3/2$ is plotted in the top panels, while $\alpha=1$ is plotted 
in the bottom panels.
By construction, the gas and the dark matter density profiles 
are very similar to each other in the outer parts of the haloes.

The gas density profile is always shallower than the dark matter 
density profile in the central part of the halo, developing 
an approximate core. 
This is because the large thermal gas pressure in the centre
balances the force of gravity exerted on the gas fluid element 
even for a nearly uniform gas density. 
The predicted gas profiles agree with numerical simulations very well:
the simulated gas profiles are also smoother in the centre
\cite{ENF98,Santa99,PTCE00,Lewis00,YJS00}.

%%%%%%%%%%%%%%%%%%%%%%%%%%%%%%%%%%%%%%%%%%%%%%%%%%%%%%%%%%%%%%%%%%%%%%
%%%%%%%%%%%%%%%%%%%%%%%%%%%%%%%%%%%%%%%%%%%%%%%%%%%%%%%%%%%%%%%%%%%%%%

\begin{figure}
  \begin{center}
    \leavevmode\epsfxsize=8.4cm \epsfbox{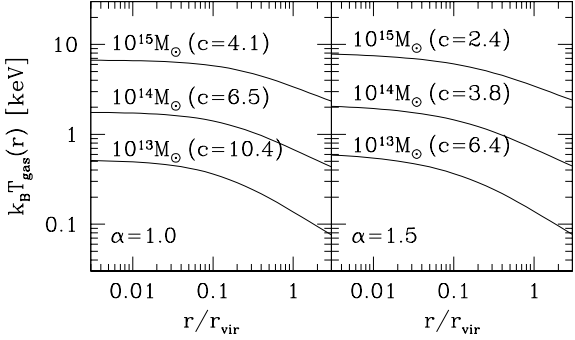}
  \end{center}
 \vspace{-1.8cm}
\caption{
 Predicted gas temperature profiles.
 The left panel uses the dark matter density profile for $\alpha=1$,
 while the right panel uses the one for $\alpha=3/2$. 
 From top to bottom lines, $M=10^{15}$, $10^{14}$, and $10^{13}M_{\sun}$
 are shown. 
 The corresponding concentration parameter, $c$, to the mass is quoted.}
\label{fig:temprofiles}
\end{figure}

Figure~\ref{fig:temprofiles} plots the gas temperature profiles
for $M=10^{15}$, $10^{14}$, and $10^{13}M_{\sun}$ from top to bottom 
lines.
The left panel shows $\alpha=1$, while the right panel shows $\alpha=3/2$.
The predicted temperature profiles also agree with the simulated
temperature profiles very well 
\cite{nfw95,ENF98,BN98,Santa99,PTCE00,YJS00,Thomas01}.

%%%%%%%%%%%%%%%%%%%%%%%%%%%%%%%%%%%%%%%%%%%%%%%%%%%%%%%%%%%%%%%%%%%%%%
%%%%%%%%%%%%%%%%%%%%%%%%%%%%%%%%%%%%%%%%%%%%%%%%%%%%%%%%%%%%%%%%%%%%%%

The predicted gas density profile for $M=10^{15}M_{\sun}$ is
remarkably similar to the simulated cluster by Frenk et al. 
\shortcite{Santa99} which has $M=1.1\times 10^{15}M_{\sun}$.
The predicted temperature profile also agrees with the simulated one;
both show the gas temperature declining with radius, which is described by 
$\gamma=1.1-1.2$. 
Using a larger sample simulated by Thomas et al. \shortcite{Thomas01},
we find the best-fitting value around $\gamma=1.1$, which agrees with our 
prediction. 

The simulations discussed above do not include radiative processes.
One may worry that including cooling and star formation may 
change the results and make the agreement worse. 
Lewis et al. \shortcite{Lewis00} have investigated this point, and
conclude that the addition of these processes modifies not only 
the central properties, but also the global structure of the haloes to 
some extent.
Nevertheless, they find that $\gamma=1.1-1.2$ gives good
fit to their simulated results outside the core 
and inside the virial radius, regardless of the inclusion of the cooling
effect. 

We conclude that our predicted profiles with the polytropic index fixed
by equation (\ref{eq:bestgamma}) fit both the observations and 
hydrodynamic simulations very well, outside the core. 
Recent {\it Chandra} 
(McNamara et al. 2000; Allen, Ettori \& Fabian 2001; Allen et al. 2001)   
and {\it XMM--Newton} (Arnaud et al. 2001a, 2001b; Kaastra et al. 2001) 
observations indicate that 
inside the core the gas temperature rises with radii.
Our model with $\gamma \sim 1.15$ cannot describe this; thus,  
we will not attempt to extend our model to the region inside the 
core.

%%%%%%%%%%%%%%%%%%%%%%%%%%%%%%%%%%%%%%%%%%%%%%%%%%%%%%%%%%%%%%%%%%%%%%
\section{Comparison with observed X-ray surface brightness profiles}

\subsection{Spherical $\beta$ profile}

Historically, X-ray observations have  
been interpreted with the spherical $\beta$ profile,
%%%%%%%%%%%%%%%%%%%%%%%%%%%%%%%%%%%%%%%%%%%%%%%%%%%%%%%%%%%%%%%%%%%
\begin{equation}
 \label{eq:betamodel}
  I_{\rm X}(r)
  =
  I_{\rm X}(0)
  \left[
   1+\left(\frac{r}{r_{\rm c}^{\rm X}}\right)^2
 \right]^{1/2-3\beta_{\rm X}},
\end{equation}
%%%%%%%%%%%%%%%%%%%%%%%%%%%%%%%%%%%%%%%%%%%%%%%%%%%%%%%%%%%%%%%%%%%
where $r_{\rm c}^{\rm X}$ is the projected X-ray core radius.
This profile had a remarkable success in fitting observed X-ray surface 
brightness profiles, except for the very central parts of haloes, where 
the radiative cooling effect seems to play a major role.
Observers often exclude this region, when they fit the measured 
X-ray profile to the $\beta$ profile.

\begin{figure}
  \begin{center}
    \leavevmode\epsfxsize=8.4cm \epsfbox{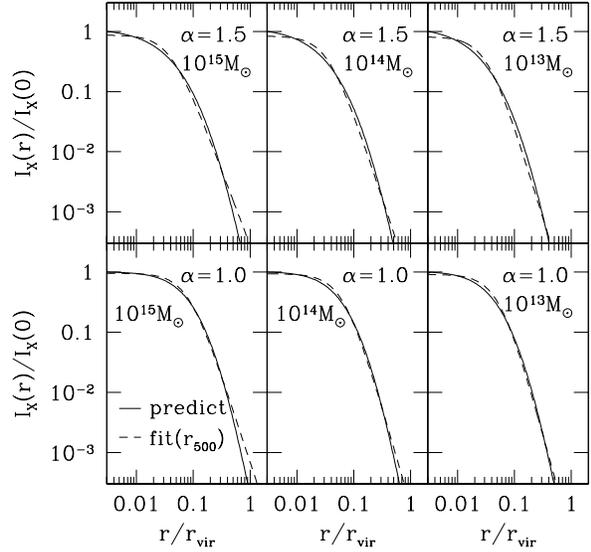}
  \end{center}
\caption{
 Predicted X-ray surface brightness profiles (solid lines), and 
 the best-fitting spherical $\beta$ profiles (dashed lines). 
 The fit is performed out to $r_{500}$. 
 Note that $r_{500}$ is half the virial radius 
 (see figure~\ref{fig:M500_Mvir}).
 The top panels use the dark matter density profile for $\alpha=3/2$,
 while the bottom panels use the one for $\alpha=1$. 
 From left to right panels, each panel shows the case of 
 $M=10^{15}$, $10^{14}$, and $10^{13}M_{\sun}$,
 respectively.}
\label{fig:xrayprof}
\end{figure}

To compare our predicted profiles with observationally 
measured X-ray profiles, we fit the core radius 
and $\beta_{\rm X}$ to our predicted X-ray surface brightness profile,
%%%%%%%%%%%%%%%%%%%%%%%%%%%%%%%%%%%%%%%%%%%%%%%%%%%%%%%%%%%%%%%%%%%
\begin{eqnarray}
  \nonumber
  I_{\rm X}(x)
  &\propto&
  \int_{-\infty}^\infty dl~\rho^2_{\rm gas}\Lambda(T_{\rm gas})\\
 \label{eq:Ix}
  &\propto&
  \int_{-\infty}^\infty dl
  \left[y_{\rm gas}\left(\sqrt{x^2+l^2}\right)\right]^{(\gamma+3)/2}.
\end{eqnarray}
%%%%%%%%%%%%%%%%%%%%%%%%%%%%%%%%%%%%%%%%%%%%%%%%%%%%%%%%%%%%%%%%%%%
Figure~\ref{fig:xrayprof} plots the predicted X-ray profiles, and 
the best-fitting $\beta$ profiles.
We find that the predicted X-ray profile is quite similar to the 
$\beta$ profile \cite{MSS98,SSM98}. 

%%%%%%%%%%%%%%%%%%%%%%%%%%%%%%%%%%%%%%%%%%%%%%%%%%%%%%%%%%%%%%%%%%%%%%
%%%%%%%%%%%%%%%%%%%%%%%%%%%%%%%%%%%%%%%%%%%%%%%%%%%%%%%%%%%%%%%%%%%%%%

\subsection{Examining observational selection effects}

The main issue in comparing quantitatively the predicted X-ray profiles with
the observed ones is out to which radius the observed 
X-ray profile is typically fitted to the $\beta$ profile. 
Our predicted profile has a slope decreasing from 
0 to $-3$ continuously with radius, corresponding to 
$\beta_{\rm X}$ varying from 0 to 1; thus, the larger outermost 
radius we use for the fit, the larger $\beta_{\rm X}$ we obtain.

\begin{figure}
  \begin{center}
    \leavevmode\epsfxsize=8.4cm \epsfbox{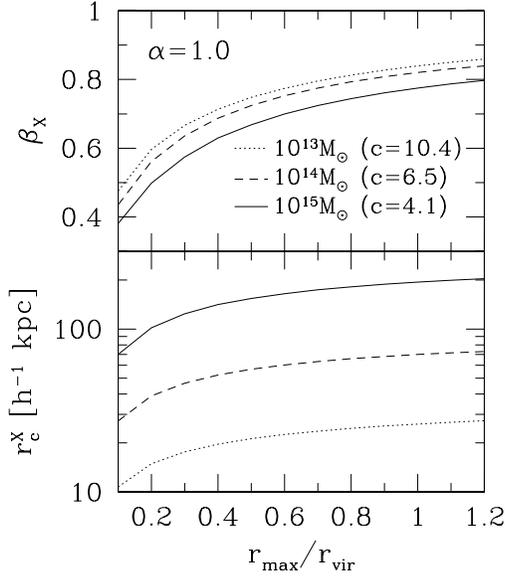}
  \end{center}
\caption{
 Sensitivity of the parameters of the spherical $\beta$ profile
 to the maximum radius, $r_{\rm max}$, used for fitting 
 the X-ray surface brightness profile to the $\beta$ profile. 
 The maximum radius is in units of the virial radius.
 The top panel shows the X-ray outer slope, $\beta_{\rm X}$, while the bottom 
 panel shows the X-ray core radius, $r_{\rm c}^{\rm X}$.
 The solid lines, dashed lines, and dotted lines indicate the virial mass 
 of $10^{15}$, $10^{14}$, and $10^{13}~M_{\sun}$, respectively.
 The corresponding concentration parameter, $c$, to the mass is quoted.
 The dark matter density profile for $\alpha=1$ is used.}
\label{fig:beta_rmax}
\end{figure}

The top panel of figure~\ref{fig:beta_rmax} shows how sensitive 
$\beta_{\rm X}$ is to the maximum radius, $r_{\rm max}$, that is used
for the fit. 
We find that $\beta_{\rm X}$ increases with $r_{\rm max}$.
Navarro et al. \shortcite{nfw95} also observe this effect 
in their hydrodynamic simulations.
We find that the sensitivity of $\beta_{\rm X}$ to $r_{\rm max}$
is comparable to the mass dependence of $\beta_{\rm X}$.
In other words, we can create any correlation between $\beta_{\rm X}$ and mass
by changing $r_{\rm max}$ slightly but systematically.
The bottom panel of the figure plots the sensitivity of the core radius 
to $r_{\rm max}$. 
We find that $r_{\rm c}^{\rm X}$ also increases with $r_{\rm max}$.

%%%%%%%%%%%%%%%%%%%%%%%%%%%%%%%%%%%%%%%%%%%%%%%%%%%%%%%%%%%%%%%%%%%%%%
%%%%%%%%%%%%%%%%%%%%%%%%%%%%%%%%%%%%%%%%%%%%%%%%%%%%%%%%%%%%%%%%%%%%%%

To make a meaningful comparison between our predicted X-ray 
profiles and observations, we have to characterize 
the observational selection effect on $r_{\rm max}$.
In X-ray observations, $r_{\rm max}$ is the radius at which the 
background noise starts to dominate the X-ray surface brightness. 
Hence, we roughly estimate $r_{\rm max}$ by equating 
$I_{\rm X}(r_{\rm max})$ to the background noise level, $BG$,
%%%%%%%%%%%%%%%%%%%%%%%%%%%%%%%%%%%%%%%%%%%%%%%%%%%%%%%%%%%%%%%%%%%%%%
\begin{equation}
 BG = I_{\rm X}(r_{\rm max})
  \propto \rho_{\rm gas}^2(0)T_{\rm gas}^{1/2}
  \left(\frac{r_{\rm max}}{r_{\rm s}}\right)^{2s_*+1}r_{\rm s},
\end{equation}
%%%%%%%%%%%%%%%%%%%%%%%%%%%%%%%%%%%%%%%%%%%%%%%%%%%%%%%%%%%%%%%%%%%%%%
where $s_*\sim -3$ is a slope of the gas density profile given by
equation (\ref{eq:sstar}) with $x_*=c$. 
As $c\propto T_{\rm gas}^{-3/10}$, 
$\rho_{\rm gas}(0)\propto c^3\propto T_{\rm gas}^{-9/10}$, 
and $r_{\rm s}=r_{\rm vir}/c\propto T_{\rm gas}^{4/5}$.
After some algebra, we obtain $r_{\rm max}$ in units of the virial radius,
$r_{\rm vir}$, 
%%%%%%%%%%%%%%%%%%%%%%%%%%%%%%%%%%%%%%%%%%%%%%%%%%%%%%%%%%%%%%%%%%%%%%
\begin{equation}
 \label{eq:selection}
  \frac{r_{\rm max}}{r_{\rm vir}}
  \propto
  T_{\rm gas}^{3/10+1/(4s_*+2)}(BG)^{1/(2s_*+1)}.
\end{equation}
%%%%%%%%%%%%%%%%%%%%%%%%%%%%%%%%%%%%%%%%%%%%%%%%%%%%%%%%%%%%%%%%%%%%%%
If we assume $BG$ not to vary significantly from observation to 
observation, we obtain $r_{\rm max}/r_{\rm vir}\propto 
T_{\rm gas}^{3/10+1/(4s_*+2)}\approx T_{\rm X}^{1/5}$.

\begin{figure}
  \begin{center}
    \leavevmode\epsfxsize=8.4cm \epsfbox{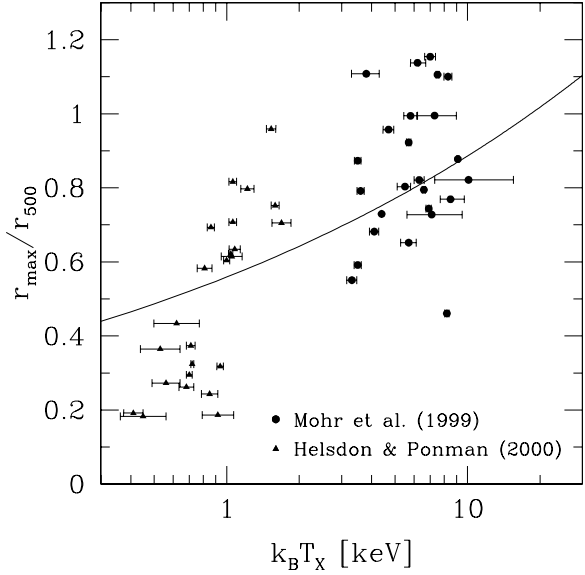}
  \end{center}
\caption{
 Observational maximum radii, $r_{\rm max}$, used for fitting
 the X-ray surface brightness profiles to the $\beta$ profile,
 as a function of the emission-weighted mean temperature, $T_{\rm X}$.
 $r_{\rm max}$ is in units of $r_{500}$.
 The circles are the data taken from Mohr et al. \shortcite{MME99},
 while the triangles are from Helsdon \& Ponman \shortcite{HP00}.
 The solid line plots an estimate of the temperature dependence of the 
 selection effect on $r_{\rm max}$ caused by the background noise 
 (equation (\ref{eq:selection})).
 Note that $r_{500}$ is half the virial radius 
 (see figure~\ref{fig:M500_Mvir}).}
\label{fig:Tx_rmax}
\end{figure}

Figure~\ref{fig:Tx_rmax} plots $r_{\rm max}$ in units of $r_{500}$ 
for observed clusters and groups.
We take the data from Mohr, Mathiesen \& Evrard \shortcite{MME99} and 
Helsdon \& Ponman \shortcite{HP00}.
We calculate $r_{500}$ from the emission-weighted mean temperature,
using the normalization of Mohr et al. \shortcite{MME99},
%%%%%%%%%%%%%%%%%%%%%%%%%%%%%%%%%%%%%%%%%%%%%%%%%%%%%%%%%%%%%%%%%%%%%%
\begin{equation}
  r_{500}= 2.37\left(\frac{T_{\rm X}}{10~{\rm keV}}\right)^{1/2}
  \left(\frac{h}{0.5}\right)^{-1}~{\rm Mpc}.
\end{equation}
%%%%%%%%%%%%%%%%%%%%%%%%%%%%%%%%%%%%%%%%%%%%%%%%%%%%%%%%%%%%%%%%%%%%%%

Despite the large scatter, there is some evidence that $r_{\rm max}/r_{500}$
increases with $T_{\rm X}$.
The figure also plots (solid line) an estimated scaling of the 
selection effect on $r_{\rm max}$ with $T_{\rm X}$ (Eq.(\ref{eq:selection})).
We have normalized it so as to have $r_{\rm max}=0.8 r_{500}$ at 6~keV,
which agrees with the observational data.
Equation~(\ref{eq:selection}) seems to capture a rough tendency
of the actual selection effect, while the large scatter around the solid 
line in the figure would be due to the background noise level,
$BG$, varying substantially from observation to observation (or 
clusters being not universal on the individual basis).

We summarize the selection effect as follows:
{\it For the cooler haloes, the smaller $r_{\rm max}/r_{500}$ is 
used for fitting the observed X-ray surface brightness profiles 
to the $\beta$ profile}.
This effect leads us to have smaller $\beta_{\rm X}$ and smaller core radii
for the cooler haloes than we have for the hotter haloes, 
according to our predictions.
We see below more quantitatively how the selection effect affects 
the comparison.

%%%%%%%%%%%%%%%%%%%%%%%%%%%%%%%%%%%%%%%%%%%%%%%%%%%%%%%%%%%%%%%%%%%%%%
%%%%%%%%%%%%%%%%%%%%%%%%%%%%%%%%%%%%%%%%%%%%%%%%%%%%%%%%%%%%%%%%%%%%%%

\subsection{X-ray core radii}

\begin{figure}
  \begin{center}
    \leavevmode\epsfxsize=8.4cm \epsfbox{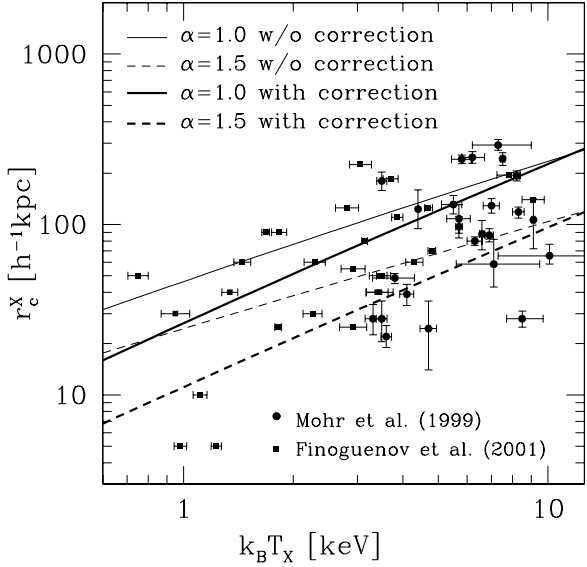}
  \end{center}
\caption{
 Predicted X-ray core radii, $r_{\rm c}^{\rm X}$, as a function of the 
 emission-weighted mean temperature, $T_{\rm X}$, in comparison with the 
 observational data from Mohr et al. \shortcite{MME99} (circles)
 and Finoguenov et al. \shortcite{FRB01} (squares).
 The solid lines represent $\alpha=1$, while the dashed lines represent 
 $\alpha=3/2$. 
 The thin lines are calculated with the predicted X-ray surface 
 brightness profiles fitted to the $\beta$ model out to $r_{500}$.
 The thick lines take into account the observational selection effect
 on the maximum radius used for the fitting
 (see figure~\ref{fig:Tx_rmax} or equation (\ref{eq:selection})).}
\label{fig:Tx_rc}
\end{figure}

Figure~\ref{fig:Tx_rc} compares our predicted X-ray core radii, 
$r_{\rm c}^{\rm X}$, with the observational data from 
Mohr et al. \shortcite{MME99} and Finoguenov et al. \shortcite{FRB01}.
We have chosen the samples from Mohr et al. \shortcite{MME99} to which 
the double-$\beta$ analysis is not applied.   
The thick lines plot the predictions corrected for the selection effect 
due to the background noise (equation~(\ref{eq:selection})), while 
the thin lines plot the predictions not corrected. 
For the latter, we have used $r_{\rm max}=r_{500}$ 
regardless of the temperature.
The selection-effect correction makes $r_{\rm max}$ smaller, 
and thus makes the core radii smaller than those without the 
correction, as the gas density profile is shallower in the inner region 
than in the outer region ($r_{\rm x}^{\rm X}$ decreases as $r_{\rm max}$
decreases; see figure~\ref{fig:beta_rmax}).

We find that the predicted core radii from our model agree
with the observational data.
The predicted core radii for $\alpha=3/2$ are systematically smaller than 
those for $\alpha=1$. 
Qualitatively, this is what one would expect, as the former profiles are 
more centrally concentrated than the latter ones at small radii (the 
rescaling of concentration with 1.7 only 
brings the outer profile in agreement, while in the centre 
differences remain). 
We also see this trend in figure~\ref{fig:xrayprof}, 
the predicted X-ray profiles.
Both the models ($\alpha=1$ and 3/2) give acceptable fit to the observations,
given the large scatter of the data. 

We have tried other concentration parameters.
We find that the concentration parameter affects the core radii less 
than the inner slope does; thus, inner gas profiles are a potentially powerful 
probe of dark matter distribution at small distances. 
Our polytropic assumption may, however, be no longer valid in the
central region, so that we need to generalize it to account for the 
observed decrease of the temperature towards the centre
(McNamara et al. 2000; Allen et al. 2001a, 2001b; 
Arnaud et al. 2001a, 2001b; Kaastra et al. 2001). 
We plan to investigate this in the future.

%%%%%%%%%%%%%%%%%%%%%%%%%%%%%%%%%%%%%%%%%%%%%%%%%%%%%%%%%%%%%%%%%%%%%%
%%%%%%%%%%%%%%%%%%%%%%%%%%%%%%%%%%%%%%%%%%%%%%%%%%%%%%%%%%%%%%%%%%%%%%

Figure~\ref{fig:Tx_rc} shows the dependence of the X-ray core radii 
on the emission-weighted mean temperature. 
The positive correlation seen in the figure is due to the 
hotter haloes being larger. 

\begin{figure}
  \begin{center}
    \leavevmode\epsfxsize=8.4cm \epsfbox{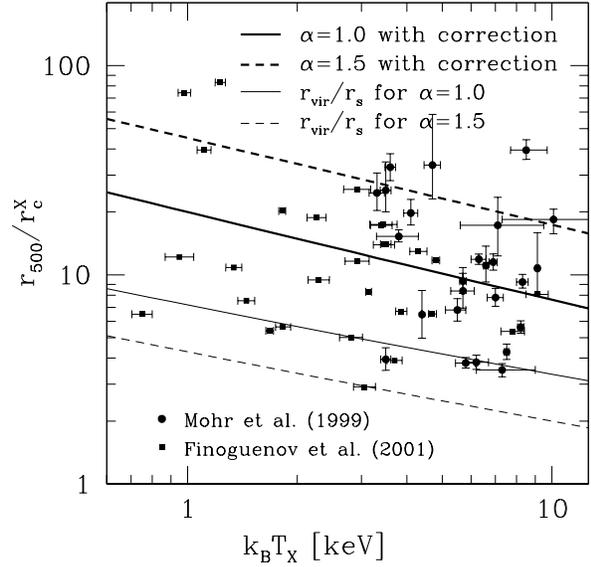}
  \end{center}
\caption{
 Predicted ratio of overdensity radii, $r_{500}$, to X-ray core radii, 
 $r_{\rm c}^{\rm X}$, as a function of the emission-weighted mean 
 temperature, $T_{\rm X}$, in comparison with the observational data from 
 Mohr et al. \shortcite{MME99} (circles) and Finoguenov et al. 
 \shortcite{FRB01} (squares).
 The solid lines represent $\alpha=1$, while the dashed lines represent 
 $\alpha=3/2$. 
 The thick lines plot $r_{500}/r_{\rm c}^{\rm X}$ with taking into 
 account the observational selection effect on the maximum radius used for 
 the fitting (see figure~\ref{fig:Tx_rmax} or equation (\ref{eq:selection})).
 The thin solid lines show the concentration parameter, 
 $c\equiv r_{\rm vir}/r_{\rm s}$.}
\label{fig:Tx_c}
\end{figure}

To investigate the concentration dependence on the mass or 
temperature, we plot in figure~\ref{fig:Tx_c} 
$r_{500}/r_{\rm c}^{\rm X}$ as a function of the emission-weighted 
mean temperature.
Our predictions for $\alpha=1$ agree with the data.
There is a marginal tendency for cooler haloes to have a higher 
$r_{500}/r_{\rm c}^{\rm X}$, and hence a higher concentration.
We also plot the concentration parameter (thin lines) in the figure;
it shows the same correlation as $r_{500}/r_{\rm c}^{\rm X}$, 
supporting our interpretation. 

We find $r_{500}/r_{\rm c}^{\rm X}\sim 0.4c$ for $\alpha=1$, and 
$r_{500}/r_{\rm c}^{\rm X}\sim 0.1c$ for $\alpha=1.5$.
The former gives $r_{\rm c}^{\rm X}\sim 0.2r_{\rm s}$, while the latter gives 
$r_{\rm c}^{\rm X}\sim 0.05r_{\rm s}$.
The former is similar to Makino et al. \shortcite{MSS98} despite
our model not assuming the isothermality.
Note that their predicted core radii are systematically smaller than
the observational data, because the concentration parameter they use 
is larger than that we use on cluster mass scales.

An important observation from figure~\ref{fig:Tx_rc} is that the data 
show fairly large scatter in $r_{500}/r_{\rm c}^{\rm X}$. 
While this could be explained to some extent 
by the uncertainties in the observational 
selection effect, it could also imply that the universal gas 
profile as proposed in this paper can only be valid in a statistical sense, 
and thus individual clusters may deviate from it significantly.
This is consistent with the findings from numerical simulations 
{\cite{Thomas01}}, but has not been shown with the observational 
data on cluster mass scales.

%%%%%%%%%%%%%%%%%%%%%%%%%%%%%%%%%%%%%%%%%%%%%%%%%%%%%%%%%%%%%%%%%%%%%%
%%%%%%%%%%%%%%%%%%%%%%%%%%%%%%%%%%%%%%%%%%%%%%%%%%%%%%%%%%%%%%%%%%%%%%

\subsection{X-ray outer slope $\beta_{\rm X}$}

\begin{figure}
  \begin{center}
    \leavevmode\epsfxsize=8.4cm \epsfbox{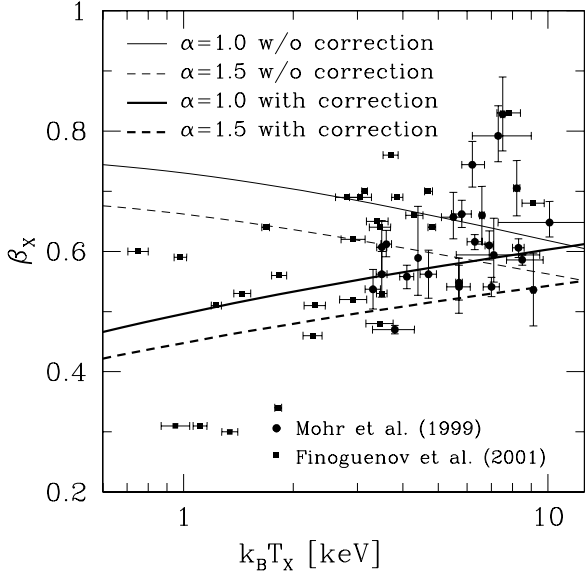}
  \end{center}
\caption{
 Predicted X-ray outer slope, $\beta_{\rm X}$, as a function of 
 the emission-weighted mean temperature, $T_{\rm X}$, in comparison with 
 the observational data from Mohr et al. \shortcite{MME99} (circles)
 and Finoguenov et al. \shortcite{FRB01}.
 The solid lines represent $\alpha=1$, while the dashed lines represent 
 $\alpha=3/2$. 
 The thin lines are calculated with the predicted 
 X-ray surface brightness profile fitted to the $\beta$ model out to $r_{500}$.
 The thick lines take into account the observational selection effect
 on the maximum radius used for the fitting 
 (see figure~\ref{fig:Tx_rmax} or equation (\ref{eq:selection})).}
\label{fig:Tx_beta}
\end{figure}

Figure~\ref{fig:Tx_beta} plots the X-ray outer slope, $\beta_{\rm X}$, 
as a function of the emission-weighted mean temperature, $T_{\rm X}$. 
The observational data show a positive correlation between 
$\beta_{\rm X}$ and $T_{\rm X}$, which is consistent with what 
Horner, Mushotzky \& Scharf \shortcite{HMS99} and 
Lloyd-Davies, Ponman \& Cannon \shortcite{LPC00} find 
in a different observational catalogue. 
Similarly, Neumann \& Arnaud \shortcite{NA99} find a positive 
correlation between $\beta_{\rm X}$ and X-ray core radii.
Since the X-ray core radius increases with $T_{\rm X}$ 
(see figure~\ref{fig:Tx_rc}), 
it follows that $\beta_{\rm X}$ also increases with $T_{\rm X}$.

%%%%%%%%%%%%%%%%%%%%%%%%%%%%%%%%%%%%%%%%%%%%%%%%%%%%%%%%%%%%%%%%%%%%%%
%%%%%%%%%%%%%%%%%%%%%%%%%%%%%%%%%%%%%%%%%%%%%%%%%%%%%%%%%%%%%%%%%%%%%%

In contrast, our model predicts a {\it negative} correlation 
between $\beta_{\rm X}$ and $T_{\rm X}$, 
if we fit the X-ray profile to the $\beta$ profile out to 
a {\it fixed} overdensity radius such as $r_{500}$.
This is easily understood: the smaller the mass is, the more
centrally concentrated the profile becomes, and the 
larger $\beta_{\rm X}$ we obtain.
We must, however, also take into account the surface-brightness-cutoff
selection effect.
Once we take it into account, the predicted $\beta_{\rm X}$ 
comes to agree with the data very well.
We show this explicitly in figure~\ref{fig:Tx_beta}.
The selection effect makes the maximum radii used for the fit smaller 
for cooler haloes, and thus makes $\beta_{\rm X}$ smaller
($\beta_{\rm X}$ decreases as $r_{\rm max}$ decreases; 
see figure~\ref{fig:beta_rmax}).

The selection effect would particularly be prominent for the 
group sample ($T_{\rm X}\sim 0.4-2$~keV) 
of Helsdon \& Ponman \shortcite{HP00}, 
for which $r_{\rm max}$ is much smaller than $r_{500}$. 
They find $\beta_{\rm X} \sim 0.3$ for the group sample \cite{HP00},
much smaller than $\beta_{\rm X}\sim 0.6$ for clusters.

We conclude that the observational selection effect causes
the positive correlation between $\beta_{\rm X}$ and $T_{\rm X}$.
We predict that the correlation becomes negative, if we fit X-ray 
profiles to the $\beta$ profile out to a fixed overdensity radius.

%%%%%%%%%%%%%%%%%%%%%%%%%%%%%%%%%%%%%%%%%%%%%%%%%%%%%%%%%%%%%%%%%%%%%%
\section{Surface brightness profiles of the Sunyaev--Zel'dovich effect}

The Sunyaev--Zel'dovich (SZ) effect \cite{ZS69} is now established 
as a very powerful observational tool for imaging clusters
(e.g., Carlstrom et al. 2000).
Measurement of surface brightness profiles of the SZ effect 
gives projected gas pressure of clusters.
To extract the gas density profile from the measurement,
we need to deproject the measured SZ surface brightness profile;
thus, we need an appropriate model to parameterize the gas density
and the gas temperature profiles.  

As often done in X-ray observations, the surface brightness profile of 
the SZ effect has also been parametrized with the $\beta$ profile,
%%%%%%%%%%%%%%%%%%%%%%%%%%%%%%%%%%%%%%%%%%%%%%%%%%%%%%%%%%%%%%%%%%%
\begin{equation}
 \label{eq:szbetamodel}
  I_{\rm SZ}(r)
  =
  I_{\rm SZ}(0)
  \left[
   1+\left(\frac{r}{r_{\rm c}^{\rm SZ}}\right)^2
 \right]^{1/2-3\beta_{\rm SZ}/2}.
\end{equation}
%%%%%%%%%%%%%%%%%%%%%%%%%%%%%%%%%%%%%%%%%%%%%%%%%%%%%%%%%%%%%%%%%%%
We compare our predicted SZ profiles with the $\beta$ profile. 
We calculate the SZ surface brightness profile from
%%%%%%%%%%%%%%%%%%%%%%%%%%%%%%%%%%%%%%%%%%%%%%%%%%%%%%%%%%%%%%%%%%%
\begin{equation}
 \label{eq:Isz}
  I_{\rm SZ}(x)
  \propto
  \int_{-\infty}^\infty dl~P_{\rm gas}
  \propto
  \int_{-\infty}^\infty dl
  \left[y_{\rm gas}\left(\sqrt{x^2+l^2}\right)\right]^{\gamma}.
\end{equation}
%%%%%%%%%%%%%%%%%%%%%%%%%%%%%%%%%%%%%%%%%%%%%%%%%%%%%%%%%%%%%%%%%%%

\begin{figure}
  \begin{center}
    \leavevmode\epsfxsize=8.4cm \epsfbox{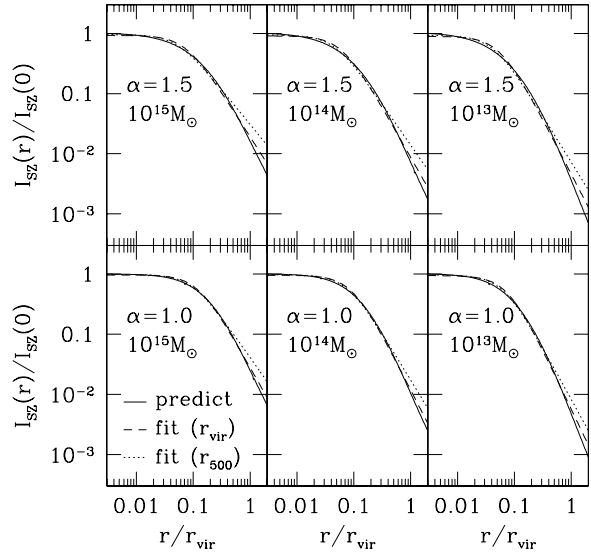}
  \end{center}
\caption{
 Predicted surface brightness profiles of the Sunyaev--Zel'dovich effect
 (solid lines), and the best-fitting spherical $\beta$ profiles 
 (dashed and dotted lines). 
 The fit is performed out to either the virial radius (dashed lines) 
 or $r_{500}$ (dotted lines). 
 Note that $r_{500}$ is half the virial radius 
 (see figure~\ref{fig:M500_Mvir}).
 The top panels use the dark matter density profile for $\alpha=3/2$,
 while the bottom panels use the one for $\alpha=1$. 
 From left to right panels, each panel shows the case of 
 $M=10^{15}$, $10^{14}$, and $10^{13}~M_{\sun}$,
 respectively.}
\label{fig:szprof}
\end{figure}

Figure~\ref{fig:szprof} shows the predicted SZ surface brightness
profiles together with the best-fitting $\beta$ profiles.
The fit is performed out to either the virial radius (dashed lines), or 
$r_{500}$ (dotted lines) at which the dark matter density 
 is 500 times the critical density of the universe. 
The SZ profiles are substantially shallower than the X-ray 
profiles (compare this figure with figure~\ref{fig:xrayprof}).
This comes from the different weights in the gas density for 
the SZ and X-ray profiles: the SZ surface brightness is weighted 
by $\rho_{\rm gas}$, while the X-ray surface brightness is weighted 
by $\rho_{\rm gas}^2$.

Although the $\beta$ profile gives a reasonable fit to our profile,
the slope of the fitted profile is quite sensitive to the outermost 
radius used for the fit. 
This tendency is also seen in fitting the X-ray
profiles to the $\beta$ profile, indicating that a simple power-law 
profile such as the $\beta$ profile describes
neither the X-ray nor the SZ profiles properly.

%%%%%%%%%%%%%%%%%%%%%%%%%%%%%%%%%%%%%%%%%%%%%%%%%%%%%%%%%%%%%%%%%%%%%%
%%%%%%%%%%%%%%%%%%%%%%%%%%%%%%%%%%%%%%%%%%%%%%%%%%%%%%%%%%%%%%%%%%%%%%

If the gas is isothermal, then the $\beta$ profile implies that the gas 
density profile is given by
%%%%%%%%%%%%%%%%%%%%%%%%%%%%%%%%%%%%%%%%%%%%%%%%%%%%%%%%%%%%%%%%%%%%%%
\begin{equation}
 \label{eq:isobetamodel}
  \rho_{\rm gas}(r)
  =
  \rho_{\rm gas}(0)
  \left[1+\left(\frac{r}{r_{\rm c}}\right)^2\right]^{-3\beta/2},
\end{equation}
%%%%%%%%%%%%%%%%%%%%%%%%%%%%%%%%%%%%%%%%%%%%%%%%%%%%%%%%%%%%%%%%%%%%%%
with $r_{\rm c} = r_{\rm c}^{\rm X} = r_{\rm c}^{\rm SZ}$ and
$\beta = \beta_{\rm X} = \beta_{\rm SZ}$.
As we have shown, our predicted profiles deviate
from this spherical isothermal $\beta$ model appreciably.

\begin{figure}
  \begin{center}
    \leavevmode\epsfxsize=8.4cm \epsfbox{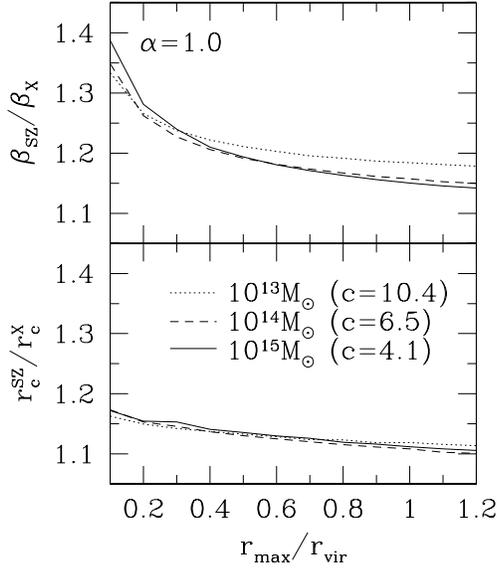}
  \end{center}
\caption{
 Comparison between fitted parameters of the $\beta$ profile obtained from 
 the X-ray and the Sunyaev--Zel'dovich surface brightness profiles.  
 The ratio of the SZ outer slope to the X-ray outer slope, 
 $\beta_{\rm SZ}/\beta_{\rm X}$, is plotted in the top panel, 
 while the ratio of the SZ core radius to the X-ray core radius, 
 $r^{\rm SZ}_{\rm c}/r^{\rm X}_{\rm c}$, is plotted in the bottom panel.
 They are shown as a function of the maximum radius, $r_{\rm max}$, 
 out to which the surface brightness profiles are fitted to the 
 $\beta$ profile. 
 The maximum radius is in units of the virial radius.
 The solid lines, dashed lines, and dotted lines indicate the virial mass 
 of $10^{15}$, $10^{14}$, and $10^{13}~M_{\sun}$, respectively.
 The corresponding concentration parameter, $c$, to the mass is quoted.
 The dark matter density profile for $\alpha=1$ is used.}
\label{fig:comp_xsz}
\end{figure}

One way to quantify {\it how much} the $\beta$ model deviates from 
our model is to calculate $\beta_{\rm SZ}/\beta_{\rm X}$ and 
$r_{\rm c}^{\rm SZ}/r_{\rm c}^{\rm X}$.
If the spherical isothermal $\beta$ model predicted exactly the same
surface brightness profiles as our model, then these ratios should be 
exactly 1.
Figure~\ref{fig:comp_xsz} shows predicted $\beta_{\rm SZ}/\beta_{\rm X}$ 
in the top panel and $r_{\rm c}^{\rm SZ}/r_{\rm c}^{\rm X}$ in the bottom
panel. 
They are plotted as a function of the maximum radius, 
$r_{\rm max}$, the outermost radius used for the fit.
We find that $\beta_{\rm SZ}/\beta_{\rm X}= 1.15-1.2$ and 
 $r_{\rm c}^{\rm SZ}/r_{\rm c}^{\rm X}= 1.1-1.15$ for 
a relevant range of $r_{\rm max}$; thus, we predict that the SZ 
profiles give systematically larger $\beta$ and $r_{\rm c}$
than the X-ray profiles (by roughly 20\%).
Yoshikawa, Itoh \& Suto \shortcite{YIS98} also observe this trend for
their simulated X-ray and SZ clusters.
The upcoming SZ observations will test this prediction, and hence
our universal gas and temperature profiles.

%%%%%%%%%%%%%%%%%%%%%%%%%%%%%%%%%%%%%%%%%%%%%%%%%%%%%%%%%%%%%%%%%%%%%%
%%%%%%%%%%%%%%%%%%%%%%%%%%%%%%%%%%%%%%%%%%%%%%%%%%%%%%%%%%%%%%%%%%%%%%

%%%%%%%%%%%%%%%%%%%%%%%%%%%%%%%%%%%%%%%%%%%%%%%%%%%%%%%%%%%%%%%%%%%%%%
\section{Mass--Temperature Relation}

Mass and emission-weighted mean temperature measured for many clusters
and groups are very tightly correlated. 
This correlation is called the mass--temperature scaling relation. 
Qualitatively, it arises from the virial relation between 
the virial mass and the virial temperature, 
$T_{\rm gas}(r_{\rm vir})\propto M_{\rm vir}^{2/3}$. 
Traditionally, workers have used a self-similar model with a constant 
concentration to explain this relation, which gives the same 
scaling as for the virial quantities.
Hydrodynamic simulations have been used to normalize the relation.

Our prediction for the mass--temperature scaling relation is
%%%%%%%%%%%%%%%%%%%%%%%%%%%%%%%%%%%%%%%%%%%%%%%%%%%%%%%%%%%%%%%%%%%
\begin{equation}
  k_{\rm B}T_{\rm X}= \eta_{\rm X}
  \frac{G\mu m_{\rm p}M_{\rm vir}}{3r_{\rm vir}},
\end{equation}
%%%%%%%%%%%%%%%%%%%%%%%%%%%%%%%%%%%%%%%%%%%%%%%%%%%%%%%%%%%%%%%%%%%
where $\eta_{\rm X}$ is the emission-weighted mean
normalization factor given by equation~(\ref{eq:eta_x}).
Figure~\ref{fig:eta_mass} plots $\eta_{\rm X}$ as a function of the 
virial mass.
We have fixed the normalization factor at the centre, $\eta(0)$, 
with equation~(\ref{eq:eta0}).
For comparison with the observational data, we use $M_{500}$ instead 
of $M_{\rm vir}$. 
We calculate $M_{500}$ from $M_{\rm vir}$ with equation~(\ref{eq:delta2vir}).

Since the concentration parameter in our model depends on the mass, 
the resulting mass--temperature relation is no longer the self-similar 
relation, $T_{\rm X}\propto M_{500}^{2/3}$. 
Moreover, the overall normalization is no longer a free parameter,
but fixed, in our model for a given universal dark matter density profile. 
We will show below that both features bring our model into  
better agreement with the data than the self-similar model. 

We first begin by discussing the mass--temperature relation for 
hotter clusters than 3~keV. 
The X-ray observations suggest that the observed emission-weighted 
mean normalization is slightly, but systematically, higher than 
predicted by hydrodynamic simulations (Horner et al. 1999; 
Nevalainen, Markevitch, \& Forman 2000; Finoguenov et al. 2001).
It means a higher emission-weighted mean temperature 
for a given mass. 
For example, Finoguenov et al. \shortcite{FRB01} find that the 
sample above 3~keV gives about 50\% higher normalization, 
$\eta_{\rm X}({\rm obs})\sim 2.0$, than the simulation by 
Evrard et al. \shortcite{EMN96} which gives 
$\eta_{\rm X}({\rm sim})\sim 1.3$; however, there is scatter among 
different hydrodynamic simulations: Bryan \& Norman \shortcite{BN98} give 
$\eta_{\rm X}({\rm sim})= (3/2)f_{\rm T}\sim 1.2$,
Eke et al. \shortcite{ENF98} give 
$\eta_{\rm X}({\rm sim})=(3/2)\overline{\beta}^{-1}_{\rm TM}\sim 1.5$,
Yoshikawa et al. \shortcite{YJS00} give 
$\eta_{\rm X}({\rm sim})=\gamma\sim 1.6$,
and Thomas et al. \shortcite{Thomas01} give $\eta_{\rm X}({\rm sim})\sim 1.5$.
Here, the second equality quotes the author's notation.
Presumably this scatter is caused by varying resolution, numerical 
techniques, methods of analyzing the simulations, and so on.

\begin{figure}
  \begin{center}
    \leavevmode\epsfxsize=8.4cm \epsfbox{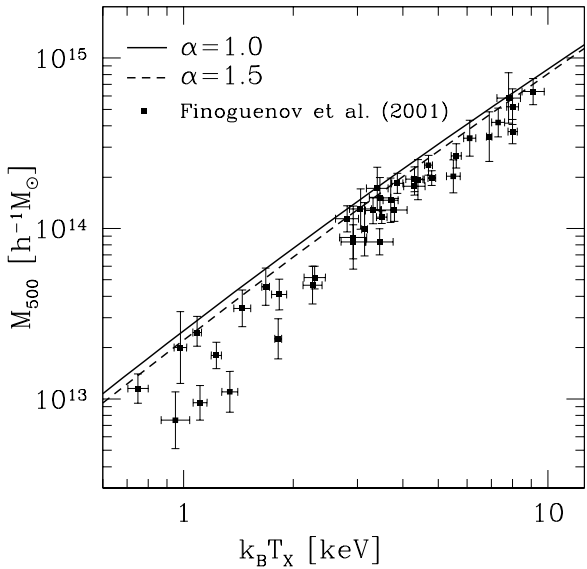}
  \end{center}
\caption{
 Predicted mass--temperature scaling relations at $z=0$
 (solid and dashed lines), compared with the observational 
 data from Finoguenov et al. \shortcite{FRB01} (squares).
 The solid line represents $\alpha=1$, while 
 the dashed line represents $\alpha=3/2$.}
\label{fig:MT}
\end{figure}

Our model predicts a normalization at the upper end of the simulations,
$\eta_{\rm X}\ga 1.5-1.8$, depending on $\alpha$ and the mass of haloes 
(see figure~\ref{fig:eta_mass}).
This means that it fits the observational data reasonably well in the 
high temperature regime.
Figure~\ref{fig:MT} compares our predictions for the mass--temperature
relation at $z=0$ (solid and dashed lines) with the local observational 
data from Finoguenov et al. \shortcite{FRB01}.
The agreement between the model and the data is reasonably good, 
especially for $k_{\rm B}T_{\rm X}>3~{\rm keV}$, 
where our model is only 20\% above the 
observations. Such small discrepancies could easily be explained 
by either systematic effects in observations or by the scatter in 
the cluster properties, which may increase the average polytropic 
index slightly (we find excellent agreement for $\gamma=1.2$).
Note that an additional kinetic pressure contribution in the hydrostatic 
equilibrium equation, $10-20$\% of the thermal pressure
contribution \cite{nfw95}, does not bring our model into better
agreement; it causes both the data and the model to underestimate 
the mass by the same amount. 

Finoguenov et al. \shortcite{FRB01} excised the central region of 
the clusters when they measured $T_{\rm X}$.
Unfortunately, it is not clear what inner cut-off radius they have used 
in the analysis.
Hence, we compare $T_{\rm X}$ for which the central region inside 
$r_{\rm cut}=0.01c$, $0.05c$, and $0.1c$ is excised, with $T_{\rm X}$
with no excision.
$0.1c$ corresponds to 340 and 150~kpc for $M_{500}=10^{15}$ and 
$10^{14}~M_{\sun}$, respectively, so it is a fairly extreme excision.
In reality, the excised region is usually smaller than the core radius
that is about $30-300$~kpc (see figure~\ref{fig:Tx_rc}); thus,
$r_{\rm cut}=0.1c$ gives a conservative upper limit on the excision effect.
We find that $r_{\rm cut}=0.1c$ reduces $T_{\rm X}$ by about 5\% and 10\% for 
$M_{500}=10^{15}~M_{\sun}$ and $10^{14}~M_{\sun}$, respectively.
The excision affects the smaller haloes more, as they are 
more centrally concentrated.
The excision also eliminates the difference between $\alpha=1$ and $1.5$
models.
The cut-off radius of below $0.05c$ has negligible effect on $T_{\rm X}$
 compared with the no excision case.
Figure~\ref{fig:MT} plots the predictions with no excision.

%%%%%%%%%%%%%%%%%%%%%%%%%%%%%%%%%%%%%%%%%%%%%%%%%%%%%%%%%%%%%%%%%%%%%%
%%%%%%%%%%%%%%%%%%%%%%%%%%%%%%%%%%%%%%%%%%%%%%%%%%%%%%%%%%%%%%%%%%%%%%

The X-ray observations of cooler ($<3~{\rm keV}$) clusters or groups
suggest that the slope of the mass--temperature scaling relation 
deviates from the virial scaling relation, $T_{\rm X}\propto M_{500}^{2/3}$
\cite{NMF00,Sato00,FRB01}.
It becomes steeper for the cooler haloes;
the cooler haloes have higher $T_{\rm X}$ than predicted by 
the virial scaling relation.
This is exactly what we predict from the model for which the 
concentration parameter decreases with the mass. 

Figure~\ref{fig:eta_mass} shows how our predicted emission-weighted 
mass--temperature normalization factor depends on the virial mass. 
The trend is in the desired direction: smaller mass
haloes have a higher normalization, and hence higher 
temperature than predicted by the virial scaling relation.
This is due to smaller mass haloes being more centrally concentrated 
than larger mass haloes.

Figure~\ref{fig:MT} shows that our predictions for cooler haloes
fit the observational data better than the predictions 
of the hydrodynamic simulations for which the dependence of the
normalization factor on the halo mass has not been observed (or studied). 
The lack of resolution in the simulations for the smaller haloes
could explain; it would not be very surprising, given how sensitive 
the emission-weighted mean temperature is to the central concentration 
of the cluster.
We have to verify this with better simulations. 

The predictions with larger $\alpha$ or larger concentration parameters
fit the data even better, although in all cases the predictions are, 
on average, still above the data. 
Within our model, a steeper concentration dependence on mass
can resolve the discrepancy, while an alternative suggestion, the 
temperature has been raised by some energy (or entropy before the
infall) injection, is also possible (Bialek, Evrard \& Mohr 2000).
A possible systematic bias in the mass determination with the hydrostatic
equilibrium equation could play a role \cite{EMN96}.

Our results show that the emission-weighted mean temperature is 
very sensitive to the central region of the haloes, and that a simple 
self-similar model cannot be used to compare the observations with the models.
As we have shown, a more realistic model beyond the self-similar model 
that includes the concentration parameter depending on mass removes 
much of the discrepancy with the data.

%%%%%%%%%%%%%%%%%%%%%%%%%%%%%%%%%%%%%%%%%%%%%%%%%%%%%%%%%%%%%%%%%%%%%%
\section{Discussion and Conclusions}

In this paper, we have constructed a model which predicts universal gas 
density and gas temperature profiles from the universal 
dark matter density profile. 
We assume that the gas density traces the dark matter 
density in the outer parts of haloes.
Many hydrodynamic simulations have verified the assumption.

This assumption, together with the hydrostatic equilibrium and spherical 
symmetry, uniquely specifies the gas density and temperature profiles 
that are related through a polytropic equation of state. 
The polytropic assumption breaks down in the inner parts of clusters 
(well inside the gas core), as shown by recent 
{\it Chandra} (McNamara et al. 2000; Allen et al. 2001a, 2001b)
and {\it XMM--Newton} (Arnaud et al. 2001a, 2001b; Kaastra et al. 2001)
temperature profiles.
For the outer parts, however, the assumption holds, and our predictions
based on the assumption agree with both simulations and observations.

Our predicted gas density and gas temperature profiles agree
with the hydrodynamic simulations both in the profile and in the 
overall amplitude.  
The predicted polytropic index is $\gamma\simeq 1.15$, 
which agrees with the observational data from Finoguenov 
et al. \shortcite{FRB01} remarkably well.

We have predicted the core radii and the outer slopes, $\beta$,
the parameters which empirically characterize surface brightness
profiles of X-ray and the Sunyaev--Zel'dovich (SZ) effect.
Fitting the predicted X-ray profiles to these parameters,
we have found that our predicted X-ray core radii and X-ray outer slopes, 
$\beta_{\rm X}$, agree with the observations,
after taking into account the observational surface-brightness 
selection effect.
The selection effect makes $\beta_{\rm X}$ increase with the
temperature, explaining the observed trend.
The X-ray core radius is less sensitive to the selection effect,
and increases with the temperature, in agreement with the observations.

We have calculated the SZ surface brightness profiles, and fitted them to 
the $\beta$ profile. 
The fitted parameters are very sensitive to the outermost radius used 
for the fit, suggesting that the $\beta$ profile is not a proper
description of the profile.
We predict that the core radii and $\beta$ fitted from the SZ profiles are
systematically larger than those fitted from the X-ray profiles, if we 
use the same $r_{\rm max}$. 
Moreover, for the SZ effect, one may use larger $r_{\rm max}$ 
than for X-ray in the fit; it further increases $\beta_{\rm SZ}$ 
compared with $\beta_{\rm X}$.

We have predicted the mass--temperature scaling relation,
and compared it with the observational data as well as the hydrodynamic
simulations. 
We have found good agreement between our predictions and the observations, 
especially for hotter clusters. 
For cooler clusters and groups, we have found that our predictions
agree with the observations better than the hydrodynamic simulations, 
which predict much lower temperature than observed. 

The disagreement between our predictions and the simulations could be 
caused by the limited resolution in the simulations, as the 
emission-weighted mean temperature is quite sensitive to the gas 
distribution in the centre. 
On the other hand, it is also possible that some physics is still 
missing in the simulations, but does not show up in our 
phenomenological description of the temperature--density relation. 
We need careful studies of the numerical resolution and additional 
physics to resolve this issue
\cite{SO98,PTCE00,Lewis00,YJS00,Muanwong01}. 

Some disagreement between our model and the observational data remains 
for the mass dependence of the concentration parameter 
obtained from pure $N$-body simulations.
This may be removed if a steeper dependence is invoked; for example, 
the baryonic cooling contraction and subsequent readjustment of the
dark matter generically increases concentration, and the effect should
be more important for smaller mass haloes for which a larger fraction 
of the gas is cooled, and transformed into stars. 
This trend is in the right direction, and agrees with the observational 
results of Sato et al. \shortcite{Sato00} who have measured very steep 
concentration dependence on the mass.
We plan to investigate this in more detail in the future.

We have not attempted to predict other scaling relations such
as the X-ray luminosity--temperature relation and the central 
entropy--temperature relation. 
To predict them, we need at least one additional uncertain parameter,
the gas mass fraction, which is a ratio of the gas mass to the dark matter
mass. 
Our method does not constrain this parameter, as this parameter does not appear
in the hydrostatic equilibrium equation.
While the gas mass fraction may be equal to the cosmic mean
baryon fraction, $\Omega_{\rm b}/\Omega_{\rm m}$, on large scales 
\cite{WNEF93}, observational evidence suggests that the gas mass 
fraction increases with mass \cite{AE99,MME99}.
This could be explained 
if some non-negligible fraction of the gas were transformed into stars or 
expelled out of the halo. 

Since the X-ray luminosity is proportional to the gas mass fraction squared,
and the central entropy is proportional to the gas mass fraction to
$-2/3$ power, the resultant predictions are significantly affected by
the uncertainty in the gas mass fraction.
In other words, we can explain the luminosity--temperature relation 
and the central entropy--temperature relation 
without invoking any preheating models, if the gas mass
fraction varies with the temperature (Bryan 2000).
Therefore, even without changing our model, we can predict these relations 
consistent with observations, by using an appropriate gas mass 
fraction depending on mass.
In addition, since the size and the density of the core region affect 
X-ray luminosity very strongly, even small modification in the profile 
leads to large change in luminosity; thus, our model does not provide
robust prediction for the X-ray luminosity.

%%%%% Acknowledgments %%%%%

E. K. acknowledges financial support from the Japan Society for
the Promotion of Sciences.
U. S. acknowledges the support of NASA and Packard and Sloan 
Foundation Fellowships. 

%%%%%%%%%%%%%%%%%%%%%%%%%%%%%%%%%%%%%%%%%%%%%%%%%%%%%%%%%%%%%%%%%%%   

%%%%%%%%%%%%%%%%%%%%%%%%%%%%%%%%%%%%%%%%%%%%%%%%%%%%%%%%%%%%%%%%%%%
\label{lastpage}

\begin{thebibliography}{}

\bibitem[\protect\citename{Allen et al. }2001a]{AEF01}
  Allen S. W., Ettori S., Fabian A. C., 2001a,
  MNRAS, 324, 877

\bibitem[\protect\citename{Allen et al. }2001b]{Allen01}
  Allen S. W. et al., 2001b,
  MNRAS, 324, 842

\bibitem[\protect\citename{Arnaud \& Evrard }1999]{AE99}
  Arnaud M., Evrard A. E., 1999, 
  MNRAS, 305, 631

\bibitem[\protect\citename{Arnaud et al. }2001a]{Arnaud01a}
  Arnaud M. et al., 2001,
  A\&A, 365, L67

\bibitem[\protect\citename{Arnaud et al. }2001b]{Arnaud01b}
  Arnaud M., Neumann D. M., Aghanim N., Gastaud R., Majerowicz S.,
Hughes J. P., 2001,
  A\&A, 365, L80

\bibitem[\protect\citename{Bialek et al. }2000]{BEM00}
  Bialek J. J., Evrard A. E., Mohr J. J.,
  ApJ, 555, 597

\bibitem[\protect\citename{Binney \& Tremaine }1987]{BT87}
  Binney J., Tremaine S., 1987, 
  Galactic Dynamics, Princeton Univ. Press, Princeton, NJ

\bibitem[\protect\citename{Bryan }2000]{Bryan00}
  Bryan G. L., 2000, 
  ApJ, 544, L1

\bibitem[\protect\citename{Bryan \& Norman }1998]{BN98}
  Bryan G. L., Norman M. L., 1998, 
  ApJ, 495, 80

\bibitem[\protect\citename{Bullock et al. }2001]{Bullock01}
  Bullock J. S., Kolatt T. S., Sigad Y., Somerville R. S., 
Kravtsov A. V., Klypin A. A., Primack J. R., Dekel A., 2001, 
  MNRAS, 321, 559

\bibitem[\protect\citename{Carlstrom et al. }2000]{Carl00}
  Carlstrom J. E., Joy M. K., Grego L., Holder G. P., Holzapfel W. L., 
Mohr J. J., Patel S., Reese E. D., 2000,
  Physica Scripta Volume T, 85, 148

\bibitem[\protect\citename{Eke et al. }1998]{ENF98}
  Eke V. R., Navarro J. F., Frenk C. S., 1998, 
  ApJ, 503, 569

\bibitem[\protect\citename{Eke et al. }2000]{ENS00}
  Eke V. R., Navarro J. F., Steinmetz, 2000, 
  ApJ, 554, 114

\bibitem[\protect\citename{Evrard et al. }1996]{EMN96} 
  Evrard A. E., Metzler C. A., Navarro J. F., 1996, 
  ApJ, 469, 494 

\bibitem[\protect\citename{Finoguenov et al. }2001]{FRB01} 
  Finoguenov A., Reiprich T. H., B\"ohringer H., 2001, 
  A\&A, 368, 749 

\bibitem[\protect\citename{Frenk et al. }1999]{Santa99} 
  Frenk C. S. et al., 1999, 
  ApJ, 525, 554

\bibitem[\protect\citename{Helsdon \& Ponman }2000]{HP00}
  Helsdon S. P., Ponman T. J.,
  MNRAS, 315, 356

\bibitem[\protect\citename{Horner et al. }1999]{HMS99} 
  Horner D. J., Mushotzky R. F., Scharf C. A., 1999, 
  ApJ, 520, 78 

\bibitem[\protect\citename{Jing \& Suto }2000]{JS00} 
  Jing Y. P., Suto Y., 2000, 
  ApJ, 529, L69

\bibitem[\protect\citename{Kaastra et al.c}2001]{Kaastra01}
  Kaastra J. S., Ferrigno C., Tamura T., Paerels F. B. S., 
Peterson J. R., Mittaz J. P. D., 2001,
  A\&A, 365, L99

\bibitem[\protect\citename{Klypin et al. }2001]{KKBP01} 
  Klypin A., Kravtsov A. V., Bullock J., Primack J., 2001, 
  ApJ, 554, 903

\bibitem[\protect\citename{Lacey \& Cole }1993]{LC93} 
  Lacey C., Cole S., 1993, 
  MNRAS, 262, 627 

\bibitem[\protect\citename{Lewis et al. }2000]{Lewis00} 
  Lewis G. F., Babul A., Katz N., Quinn T., Hernquist L., 
Weinberg D. H., 2000, 
  ApJ, 536, 623

\bibitem[\protect\citename{Lloyd-Davies et al. }2000]{LPC00}
  Lloyd-Davies E. J., Ponman T. J., Cannon D. B.,
  MNRAS, 315, 689

\bibitem[\protect\citename{Makino et al. }1998]{MSS98} 
  Makino N., Sasaki S., Suto Y., 1998, 
  ApJ, 497, 555 

\bibitem[\protect\citename{McNamara et al.}2000]{Mc00}
  McNamara B. R. et al., 2000,
  ApJ, 534, L135

\bibitem[\protect\citename{Mohr et al. }1999]{MME99}
  Mohr J. J., Mathiesen B., Evrard A. E., 1999, 
  ApJ, 517, 627 

\bibitem[\protect\citename{Moore et al. }1998]{Moore98}
  Moore B., Governato F., Quinn T., Stadel J., Lake G., 1998, 
  ApJ, 499, L5 

\bibitem[\protect\citename{Muanwong et al. }2001]{Muanwong01}
  Muanwong O., Thomas P. A., Kay S. T., Pearce F. R., 
Couchman H. M. P., 2001, 
  ApJ, 552, L27

\bibitem[\protect\citename{Nakamura \& Suto }1997]{NS97}
  Nakamura T. T., Suto Y., 1997, 
  Prog. Theor. Phys., 97, 49

\bibitem[\protect\citename{Navarro et al. }1995]{nfw95}
  Navarro J. F., Frenk C. S., White S. D. M., 1995, 
  MNRAS, 275, 720

\bibitem[\protect\citename{Navarro et al. }1996]{NFW96}
  Navarro J. F., Frenk C. S., White S. D. M., 1996, 
  ApJ, 462, 563 

\bibitem[\protect\citename{Navarro et al. }1997]{NFW97}
  Navarro J. F., Frenk C. S., White S. D. M., 1997, 
  ApJ, 490, 493 

\bibitem[\protect\citename{Neumann \& Arnaud }1999]{NA99}
  Neumann D. M., Arnaud M., 1999, 
  A\&A, 348, 711

\bibitem[\protect\citename{Netterfield et al. }2000]{Boom01}
  Netterfield B. et al., 2001, 
  ApJ, submitted (astro-ph/0104460)

\bibitem[\protect\citename{Nevalainen et al. }2000]{NMF00}
  Nevalainen J., Markevitch M., Forman W., 2000, 
  ApJ, 532, 694 

\bibitem[\protect\citename{Peebles }1980]{Peebles80}
  Peebles P. J. E., 1980, 
  The Large Scale Structure of the Universe, Princeton Univ. Press, 
Princeton, NJ

\bibitem[\protect\citename{Pearce et al. }2000]{PTCE00}
  Pearce F. R., Thomas P. A., Couchman H. M. P., Edge A. C.,
  MNRAS, 317, 1029

\bibitem[\protect\citename{Sato et al. }2000]{Sato00}
  Sato S., Akimoto F., Furuzawa A., Tawara Y., Watanabe M., 2000, 
  ApJ, 537, L73

\bibitem[\protect\citename{Seljak }2000]{Seljak00}
  Seljak U, 2000, 
  MNRAS, 318, 203

\bibitem[\protect\citename{Suginohara \& Ostriker }1998]{SO98}
  Suginohara T., Ostriker J. P., 1998,
  ApJ, 507, 16

\bibitem[\protect\citename{Suto et al. }1998]{SSM98}
  Suto Y., Sasaki S., Makino N., 1998, 
  ApJ, 509, 544 

\bibitem[\protect\citename{Thomas et al. }2001]{Thomas01}
  Thomas P. A., Muanwong O., Pearce F. R., Couchman H. M. P., 
Edge A. C., Jenkins A., Onuora L., 2001, 
  MNRAS, 324, 450

\bibitem[\protect\citename{White et al. }1993]{WNEF93}
  White S. D. M., Navarro J. F., Evrard A. E., Frenk C. S., 1993, 
  Nat, 366, 429

\bibitem[\protect\citename{Yoshikawa et al. }1998]{YIS98}
  Yoshikawa K., Itoh M., Suto Y., 1998, 
  PASJ, 50, 203

\bibitem[\protect\citename{Yoshikawa et al. }2000]{YJS00}
  Yoshikawa K., Jing Y. P., Suto Y., 2000, 
  ApJ, 535, 593

\bibitem[\protect\citename{Zel'dovich \& Sunyaev }1969]{ZS69}
  Zel'dovich Ya. B., Sunyaev R. A., 1969,
  Ap\&SS, 4, 301

\end{thebibliography}
\end{document}